\documentclass[12pt,hyper]{JHEP3}
\usepackage{graphics,psfrag}
\usepackage{amsmath,amsthm,amssymb,epsfig,euscript,array,cite,cancel}
\usepackage{cases,empheq}
\theoremstyle{definition}

\theoremstyle{remark}

\newcounter{multieqs}

\newcommand{\be}{\begin{equation}}
\newcommand{\ee}{\end{equation}}
\newcommand{\eq}[1]{(\ref{#1})}
\newcommand{\bit}{\begin{itemize}}  \newcommand{\eit}{\end{itemize}}

\newcommand{\bm}[1]{\mbox{\boldmath $#1$}}
\newcommand{\rf}[1]{(\ref{#1})}

\def\bd{\begin{document}}
\def\ed{\end{document}}
\def\nn{\nonumber}
\def\bea{\begin{eqnarray}}
\def\eea{\end{eqnarray}}
\let\bm=\bibitem

\def\la{\langle}
\def\ra{\rangle}

\def\npb#1#2#3{Nucl. Phys. {\bf{B#1}} #3 (#2)}
\def\plb#1#2#3{Phys. Lett. {\bf{#1B}} #3 (#2)}
\def\prl#1#2#3{Phys. Rev. Lett. {\bf{#1}} #3 (#2)}
\def\prd#1#2#3{Phys. Rev. {D \bf{#1}} #3 (#2)}
\def\cmp#1#2#3{Comm. Math. Phys. {\bf{#1}} #3 (#2)}
\def\cqg#1#2#3{Class. Quantum Grav. {\bf{#1}} #3 (#2)}
\def\nppsa#1#2#3{Nucl. Phys. B (Proc. Suppl.) {\bf{#1A}}#3 (#2)}
\def\ap#1#2#3{Ann. of Phys. {\bf{#1}} #3 (#2)}
\def\ijmp#1#2#3{Int. J. Mod. Phys. {\bf{A#1}} #3 (#2)}
\def\rmp#1#2#3{Rev. Mod. Phys. {\bf{#1}} #3 (#2)}
\def\mpla#1#2#3{Mod. Phys. Lett. {\bf A#1} #3 (#2)}
\def\jhep#1#2#3{J. High Energy Phys. {\bf #1} #3 (#2)}
\def\atmp#1#2#3{Adv. Theor. Math. Phys. {\bf #1} #3 (#2)}

\def\N{{\cal N}}
\def\sst{\scriptscriptstyle}
\def\thetabar{\bar\theta}
\def\Tr{{\rm Tr}}
\def\one{\mbox{1 \kern-.59em {\rm l}}}

\def\a{\alpha}      \def\da{{\dot\alpha}}  \def\dA{{\dot A}}
\def\b{\beta}       \def\db{{\dot\beta}}  
\def\g{\gamma}  \def\G{\Gamma}  \def\dc{{\dot\gamma}}  
\def\d{\delta}  \def\D{\Delta}  \def\ddt{\dot\delta}  
\def\e{\epsilon}        \def\ve{\varepsilon}  
\def\f{\phi}    \def\F{\Phi}    \def\vvf{\f}  
\def\h{\eta}  
\def\k{\kappa}  
\def\l{\lambda} \def\L{\Lambda}  
\def\m{\mu} \def\n{\nu}  
\def\o{\omega}  
\def\p{\pi} \def\P{\Pi}  
\def\r{\rho}  
\def\s{\sigma}  \def\S{\Sigma}  
\def\t{\tau}  
\def\th{\theta} \def\Th{\Theta} \def\vth{\vartheta}  
\def\X{\Xeta}  
\def\z{\zeta}  

\def\na{\nabla}  
\def\cA{{\cal A}} \def\cB{{\cal B}} \def\cC{{\cal C}}  
\def\cD{{\cal D}} \def\cE{{\cal E}} \def\cF{{\cal F}}  
\def\cG{{\cal G}} \def\cH{{\cal H}} \def\cI{{\cal I}}  
\def\cJ{{\cal J}} \def\cK{{\cal K}} \def\cL{{\cal L}}  
\def\cM{{\cal M}} \def\cN{{\cal N}} \def\cO{{\cal O}}  
\def\cP{{\cal P}} \def\cQ{{\cal Q}} \def\cR{{\cal R}}  
\def\cS{{\cal S}} \def\cT{{\cal T}} \def\cU{{\cal U}}  
\def\cV{{\cal V}} \def\cW{{\cal W}} \def\cX{{\cal X}}  
\def\cY{{\cal Y}} \def\cZ{{\cal Z}}

\def\ua{\underline{\alpha}}  
\def\uc{\underline{\phantom{\alpha}}\!\!\!\gamma}  
\def\um{\underline{\mu}}  
\def\ud{\underline\delta}  
\def\ue{\underline\epsilon}  
\def\una{\underline a}\def\unA{\underline A}  
\def\unb{\underline b}\def\unB{\underline B}  
\def\unc{\underline c}\def\unC{\underline C}  
\def\und{\underline d}\def\unD{\underline D}  
\def\une{\underline e}\def\unE{\underline E}  
\def\unf{\underline{\phantom{e}}\!\!\!\! f}\def\unF{\underline F}  
\def\unm{\underline m}\def\unM{\underline M}  
\def\unn{\underline n}\def\unN{\underline N}  
\def\unp{\underline{\phantom{a}}\!\!\! p}\def\unP{\underline P}  
\def\unq{\underline{\phantom{a}}\!\!\! q}  
\def\unQ{\underline{\phantom{A}}\!\!\!\! Q}  
\def\unH{\underline{H}}  
  
\def\As {{A \hspace{-6.4pt} \slash}\;}  
\def\bs {{b \hspace{-6.4pt} \slash}\;}  
\def\Ds {{D \hspace{-6.4pt} \slash}\;}
\def\Gts {{\Gt \hspace{-6.4pt} \slash}\;}
\def\ds {{\del \hspace{-6.4pt} \slash}\;}  
\def\ss {{\s \hspace{-6.4pt} \slash}\;}  
\def\ks {{ k \hspace{-6.4pt} \slash}\;}  
\def\ps {{p \hspace{-6.4pt} \slash}\;}   
\def\xs {{x \hspace{-6.4pt} \slash}\;}  
\def\pas {{{p_1} \hspace{-6.4pt} \slash}\;}  
\def\pbs {{{p_2} \hspace{-6.4pt} \slash}\;}   
\def\cFs {{{\cal F} \hspace{-6.4pt} \slash}\;}

\def\Ah{{\hat{A}}}  
\def\Dh{{\hat{D}}}
\def\Gh{{\hat{G}}}
\def\Fh{{\hat{F}}}
\def\Ih{{\hat{I}}} 
\def\Jh{{\hat{J}}} 
\def\Kh{{\hat{K}}}
\def\Lh{{\hat{L}}} 
\def\Ph{{\hat{P}}}
\def\Rh{{\hat{R}}}
\def\Vh{{\hat{V}}} 
\def\Xh{{\hat{X}}}
 
\def\ah{{\hat{\a}}}
\def\bh{{\hat{\b}}}
\def\gh{{\hat{\g}}}
\def\dh{{\hat{\d}}}
\def\hh{\hat{h}}
\def\uh{\hat{u}}  
\def\xh{\hat{x}}  
\def\yh{\hat{y}}  
\def\ph{\hat{p}}  
\def\xih{\hat{\xi}}  
\def\chih{\hat{\chi}}  
\def\Psih{\hat{\Psi}}

\def\psit{\tilde{\psi}}  
\def\Psit{\tilde{\Psi}}   
\def\Psibt{\tilde{\bar{Psi}}}  

\def\st{\tilde{\sigma}}  

\def\delt{\tilde{\delta}}
\def\Phit{\tilde{\Phi}}   
\def\Phitb{\overline{\tilde{Phi}}}  
\def\tht{\tilde{\th}}  
\def\lt{\tilde{\l}}
\def\chit{\tilde{\chi}}   
\def\phit{\tilde{\phi}} 

\def\At{\tilde{A}}
\def\Bt{\tilde{B}}
\def\Ct{\tilde{C}}
\def\Dt{\tilde{D}}
\def\Et{\tilde{E}}
\def\Ft{\tilde{F}}
\def\Gt{\tilde{G}}
\def\Ht{\tilde{H}}
\def\It{\tilde{I}}
\def\Jt{\tilde{J}}
\def\Qt{\tilde{Q}}  
\def\Rt{\tilde{R}}  
\def\Mt{\tilde{M }}  
\def\Nt{\tilde{N}}   
\def\St{\tilde{S}}
\def\Vt{\tilde{V}}
\def\Xt{\tilde{X}} 
\def\at{\tilde{a}}
\def\ct{\tilde{c}}
\def\dt{\tilde{d}}
\def\htt{\tilde{h}} 
\def\ft{\tilde{f}}
\def\gt{\tilde{g}}
\def\pt{\tilde{p}}  
\def\qt{\tilde{q}}  
\def\vt{\tilde{v}}  
\def\nt{\tilde{n}}  
\def\ut{\tilde{u}}  
\def\wt{\tilde{w}}  
\def\zt{\tilde{z}} 
\def\xt{\tilde{x}} 
\def\yt{\tilde{y}} 
\def\Psit{\tilde{\Psi}}
\def\vphit{\tilde{\varphi}}

\def\cHt{\tilde{\cH}}

\def\eb{\bar{\epsilon}} 
\def\delb{\bar{\partial}}  
\def\thb{\bar{\theta}}
\def\mub{\bar{\mu}}
\def\lamb{\bar{\l}}
\def\psib{\bar{\psi}}
\def\sb{\bar{\sigma}}
\def\xib{\bar{\xi}}
\def\chib{\bar{\chi}}

\def\Psib{\bar{\Psi}}
\def\Phib{\bar{\Phi}}
\def\Lamb{\bar{\Lambda}}
\def\Sb{{\overline \Sigma}}
\def\cb{\bar{c}}
\def\hb{\bar{h}}
\def\qb{\bar{q}}
\def\wb{\bar{w}}
\def\ub{\bar{u}}
\def\zb{{\bar{z}}}
\def\Hb{\bar{H}}
\def\Qb{{\bar Q}}
\def\Omegab{\overline{\Omega}}
\def\ob{\overline{\omega}}

\def\Ab{{\overline A}} \def\Bb{{\overline B}} \def\Cb{{\overline C}}  
\def\Db{{\overline D}} \def\Eb{{\overline E}} \def\Fb{{\overline F}}  
\def\Gb{{\overline G}} 
\def\Ib{{\overline I}}  
\def\Jb{{\overline J}} \def\Kb{{\overline K}} \def\Lb{{\overline L}}  
\def\Mb{{\overline M}} \def\Nb{{\overline N}} \def\Ob{{\overline O}}  
\def\Pb{{\overline P}}  \def\Rb{{\overline R}}  
 \def\Tb{{\overline T}} \def\Ub{{\overline U}}  
\def\Vb{{\overline V}} \def\Wb{{\overline W}} \def\Xb{{\overline X}}  
\def\Yb{{\overline Y}} \def\Zb{{\overline Z}}  

\def\fb{{\overline f}}
\def\gb{{\overline g}}
\def\mb{{\overline m}}
\def\lb{{\overline l}}
\def\yb{{\overline y}}
  
\def\ldel{{\overleftarrow{\del}}}
\def\rdel{{\overrightarrow{\del}}}
\def\ldeldel{{\overleftarrow{\del^2}}}
\def\rdeldel{{\overrightarrow{\del^2}}}
\def\ldelb{{\overleftarrow{\bar{\del}}}}
\def\rdelb{{\overrightarrow{\bar{\del}}}}
\def\ba{{\bf a}} 
\def\bk{{\bf k}}  
\def\bl{{\bf l}}  
\def\bp{{\bf p}}  
\def\bq{{\bf q}}  
\def\br{{\bf r}}
\def\bt{{\bf t}}
\def\bu{{\bf u}}
\def\bv{{\bf v}}
\def\bx{{\bf x}}  
\def\by{{\bf y}}  
\def\bR{{\bf R}}  
\def\bV{{\bf V}}

\def\bone{{\bf 1}}  

\def\va{{\vec a}}
\def\vk{{\vec k}}
\def\vp{{\vec p}}
\def\vq{{\vec q}}
\def\vx{{\vec x}}
\def\vy{{\vec y}}
\def\vu{{\vec u}}
\def\vv{{\vec v}}

\def\vs{{\vec \sigma}}
\def\vtau{{\vec \tau}}

\newcommand{\ov}[1]{\overrightarrow{#1}}

\def\frA{\mathfrak{A}}
\def\frB{\mathfrak{B}}
\def\frC{\mathfrak{C}}
\def\frD{\mathfrak{D}}
\def\frE{\mathfrak{E}}
\def\frF{\mathfrak{F}}
\def\frG{\mathfrak{G}}
\def\frH{\mathfrak{H}}
\def\frM{\mathfrak{M}}
\def\frN{\mathfrak{N}}
\def\frR{\mathfrak{R}}
\def\frW{\mathfrak{W}}

\def\fra{\mathfrak{a}}
\def\frb{\mathfrak{b}}
\def\frf{\mathfrak{f}}
\def\frg{\mathfrak{g}}
\def\frh{\mathfrak{h}}
\def\frl{\mathfrak{l}}
\def\frs{\mathfrak{s}}
\def\fri{\mathfrak{i}}
\def\frj{\mathfrak{j}}

\def\ma{\mathfrak{a}}
\def\mg{\mathfrak{g}}
\def\mh{\mathfrak{h}}
\def\mR{\mathfrak{R}}
\def\mN{\mathfrak{N}}

\def\d{\delta}\def\D{\Delta}\def\ddt{\dot\delta}  
  
\def\pa{\partial} \def\del{\partial}  
\def\xx{\times}  
\def\uno{\mbox{1 \kern-.59em {\rm l}}}    
  
\def\trp{^{\top}}  
\def\inv{^{-1}}  
\def\dag{{^{\dagger}}}  
\def\pr{^{\prime}}  
  
\def\rar{\rightarrow}  
\def\lar{\leftarrow}  
\def\lrar{\leftrightarrow}  
  
\newcommand{\0}{\,\!}      %this is just NOTHING!  
\def\one{1\!\!1\,\,}  
\def\im{\imath}  
\def\jm{\jmath}  
  
\newcommand{\tr}{\mbox{tr}}  
\newcommand{\slsh}[1]{/ \!\!\!\! #1}  
  
\def\vac{|0\rangle}  
\def\lvac{\langle 0|}  
  
\def\hlf{\frac{1}{2}}  
\def\ove#1{\frac{1}{#1}}  

\def\Box{\square}  
\def\CC {\mathbb{C}}
\def\FF {\mathbb{F}}
\def\RR{\mathbb{R}}
\def\NN{\mathbb{N}}  
\def\ZZ{\mathbb{Z}}  
\def\bb#1{{\bf #1}}  
\def\bcomment#1{}  
\def\bfhat#1{{\bf \hat{#1}}}  
\def\VEV#1{\left\langle #1\right\rangle}  

\newcommand{\ex}[1]{{\rm e}^{#1}} \def\ii{{\rm i}}  

\newcommand{\lrbrk}[1]{\left(#1\right)}
\newcommand{\sfrac}[2]{{\textstyle\frac{#1}{#2}}}
 
\def\stw{{\sqrt{2}}}

\def\rf {{\rm f}}
\def\ri {{\rm i}}
\def\rj {{\rm j}}
\def\rk {{\rm k}}
\def\rl {{\rm l}}
\def\rs {{\scriptscriptstyle \rm S}}
\def\rt {{\scriptscriptstyle \rm T}}

\def\rQ {{\scriptscriptstyle \rm \cQ}}
\def\rR {{\scriptscriptstyle \rm \cR}}

\def\cBb{{\cal \Bb}}
\def\cQb{{\cal \Qb}}
\def\cRb{{\cal \Rb}}
\def\cWb{{\cal \Wb}}

\def\fd {{\rm N}}
\def\afd {{\overline{\rm N}}}

\def \II {I\hspace{-.1em}I\hspace{.1em}}
\def \IIA {\mbox{\II A\hspace{.2em}}}
\def \IIB {\mbox{\II B\hspace{.2em}}}
\def \gs {g^s}
\def \ls {\lambda^s}

\def \I {{\cal I}}
\def \qs {q\hspace{-.53em}/\hspace{.15em}}
\def \ks {k\hspace{-.53em}/\hspace{.15em}}
\def \YM {{\mbox{\tiny YM}}}
\def \gym {g_{\YM}}

\def \Lc {\L_c}
\def\IR{\relax{\rm I\kern-.18em R}}
\def \id {{\bf 1}}

\def\cci{\ell}
\def\ccj{\ell'}

\def \thbb{\overline{\th\th}}
\newcommand \ol{\overline}
\def \lamb{\bar{\lambda}}
\def \vphi{\varphi}
\def \lambh{\hat{\bar{\lambda}}}
\def \lh{\hat{\lambda}}
\def \dd{\ddagger}

\def \Xd{\dot{X}}
\def \nd{\noindent \\}

\author{Chong-Sun Chu and Sheng-Lan Ko\\  
Centre for Particle Theory
and Department of Mathematical Sciences,\\ 
Durham University, Durham, DH1 3LE, UK \\
E-mail:  
\email{chong-sun.chu@durham.ac.uk}, \email{sheng-lan.ko@durham.ac.uk} }

\title{Non-abelian Action for Multiple Five-Branes with Self-Dual Tensors}

\abstract{
We construct an action for non-abelian 2-form in 
6-dimensions. Our action consists of a 
non-abelian generalization of the abelian action of Perry and Schwarz 
for a single 
five-brane.
It admits a self-duality equation on the field strength 
as the equation of motion. It has a modified 6d Lorentz symmetry. 
On dimensional reduction on a circle, our action gives the standard 5d 
Yang-Mills action
plus higher order corrections.
Based on these properties,
we propose 
that our theory describes 
the gauge sector of 
multiple M5-branes in flat space.
}
\preprint{DCPT-12/09}
\keywords{M-Theory, D-branes, M-branes, Gauge Symmetry}

\begin{document}

\section{Introduction}

The low energy theory of $N$ coincident M5-branes 
is given by an interacting (2,0) superconformal
theory in 6 dimensions \cite{zero}. 
For a single M5-brane, the low energy theory
is known \cite{howe,PS,schw1,pst,nilsson}. 
So far very little is known about this theory for $N>1$.
There are a number of difficulties associated with this theory.
First, the structure of (2,0) supersymmetry constraints the 2-form potential
to have self-dual field strength. This makes it difficult to write
down a Lorentz invariant action. This problem was solved in
\cite{PS,schw1,pst} where an action principle was constructed 
with the self-duality
equation obtained as the equation of motion. 
For the
non-abelian case, there is an additional problem that
an appropriate  generalization of the tensor gauge symmetry was not
known. 
In particular, there are no-go
theorems \cite{no-go} which state that there is no 
nontrivial deformation of the Abelian 2-form gauge theory 
if locality of the action and the 
transformation laws are assumed. The no-go theorems suggest an
important direction to go is to give up locality.

Since M2-branes can end on M5-branes, one may wonder what one may
learn by considering the intersecting M2-M5 branes system.    
In the paper \cite{CS2}, a system of open $N$ M2-branes  
described by the open ABJM theory \cite{ABJM} is considered. 
The gauge non-invariance of the boundary Chern-Simons action  was 
shown to imply 
the existence of a Kac-Moody current algebra
on the worldsheet of multiple self-dual strings. It was conjectured \cite{chu}
that  the Kac-Moody symmetry induces a $U(N)\times U(N)$ 
gauge symmetry in the theory of $N$ coincident M5-branes.
The precise nature of this gauge symmetry 
in the theory of M5-branes is however not known due to our little 
understanding of the self-dual strings.
Motivated by this, in \cite{chu}
a set of $U(N)\times U(N)$ gauge bosons was introduced 
and a version of non-abelian generalization of the tensor gauge symmetry of
2-form gauge potentials was constructed. 
%c9
This formulation has the advantage of having manifest Lorentz symmetry fully. 

Generally, the non-abelian tensor gauge symmetry is
linearly represented if the  $U(N)\times U(N)$  gauge bosons are
treated as independent fields. On the other hand, the (2,0)
supersymmetry of M5-branes implies that no extra degrees of freedom
is allowed and so these fields must be taken as auxiliary. This turns
out to be very difficult for one of the auxiliary fields. So in this
paper we will consider a 
%c10 different 
gauge fixed approach 
%c10 to construct a theory of multiple M5-branes
%c9
by given up manifest 6d Lorentz symmetry. 

%c3 
%c10
As a first step towards understanding the theory of multiple M5-branes, we will
focus on the chiral tensor gauge fields in this paper. 
Our action
consists of a  non-abelian generalization of the  action of Perry and Schwarz
\cite{PS} plus an additional term which sets the Yang-Mills gauge fields
to become auxiliary. We emphasize
% k7 emphasis -> emphasize 
 that the 
action of Perry-Schwarz (PS) is of the same type as the action
originally introduced by Henneaux and Teitelboim (HT) \cite{HT}, 
see also \cite{BH} for a recent discussion.
% k4 request by Henneaux 
The difference is that a time direction was separated from the rest 
in HT action as they were interested in a Hamiltonian
description, while in the PS action a space direction was separated from the 
(5+1) dimensional spacetime, making it particularly suitable for discussing 
dimensional reduction of the system 
\footnote{The covariant Pasti-Sorokin-Tonin (PST) 
formulation \cite{pst} unifies both since
one can gauge fix the auxiliary 
% k3 auxiliariy
scalar to arrive at these different 
formulations.}. Since we will be interested in dimensional reduction of our 
action, so we will follow \cite{PS} in this paper.   
As in Perry-Schwarz's construction, a direction
$x_5$ is singled out and specially treated, so our theory is only
manifestly 5d Lorentz invariant. Nevertheless, we manage to establish
the existence of an additional non-manifest 6d Lorentz symmetry,
generalizing the result of the abelian case \cite{PS,HT}. 
Moreover,  on dimensional reduction on a circle, our action 
gives rise directly to the standard 5d Yang-Mills theory 
%c9
plus higher order corrections.
Based on these properties, we propose that 
our action describes 
%c10
the gauge sector of a system of coincident M5-branes in flat space.
The tensor gauge symmetry in our action
%c3 
turns out to be abelian, but highly nonlinear and nonlocal.
In fact whether the tensor gauge symmetry is 
abelian or non-abelian is not constrained by any physical requirement
we know of.
The abelian nature of the tensor gauge symmetry is thus
a prediction of our construction. The construction of a non-abelian tensor
gauge symmetry is still an interesting mathematical question, but from our
construction it seems not necessary for the 
%c9
non-covariant description of multiple M5-branes.

The plan of the paper is as follows. In section 2, we  review
the construction of Perry and Schwarz \cite{PS}.
In section 3, we present our construction of the action
for non-abelian 2-form fields and establish the properties of 
self-duality, 6d Lorentz symmetry and dimensional reduction to
5d Yang-Mills action. Section 4 contains some further discussions. 
%c10
In particular we 
comment on the inclusion of fermions and
scalar fields and supersymmetry in the discussion section. 
For completeness, three
appendices are included which treat some 
analysis in the main text in more details. 

Recent related works on the subject includes:
\cite{dou,lam2} which proposed a fundamental definition of
multiple M5-branes in terms of 5d supersymmetric Yang-Mills theory;
\cite{lam1} which constructed a non-abelian version of (2,0)
supersymmetric equation of motion using Lie 3-algebra; 
\cite{ho} which constructed a compactified theory of 
non-abelian 2-form gauge potentials with a self-dual field strength;
\cite{sezgin} which proposed a more general framework than \cite{chu}
in utilizing a 3-form gauge potentials in addition to the 1-form gauge
potentials; \cite{qg,CG,qg2} which studied the form of quantum geometry of
M5-branes in a $C$-field background; \cite{amp} on amplitudes of
multiple M5-branes theory; \cite{lee} on the 
$N^3$ entropy counting of M5-branes;
as well as other issues concerning multiple M5-branes \cite{others}. 
For a review 
on older results on 
M5-branes and superconformal theory in 6-dimensions, we suggest
\cite{r1}.

%c7
\section{Abelian Action of Perry-Schwarz}

Let us start by reviewing the construction \cite{PS, HT} of an action for
a self-dual tensor  in 
6-dimensions. A key feature of their construction is that a 
certain direction, $x^0$ in \cite{HT} or $x^5$ in \cite{PS}, 
has to be singled out and so the formulation has only 
manifestly 5d rotational invariance or 5d Lorentz invariance. 
Nevertheless these theories do possess the full Lorentz symmetry.
The existence of this modified 
Lorentz symmetry is a remarkable feature of these constructions.
 
We will be interested in the Lagrangian formulation of 
%c10 
the chiral tensor gauge fields on  
multiple 
M5-branes and its dimensional reduction. Therefore let us 
follow the construction of Perry-Schwarz \cite{PS} in the following.
Let us 
denote the 5d and 6d coordinates by $x^\m= (x^0, x^1, \cdots, x^4)$ and
$x^M= (x^\m, x^5)$. We adopt the  convention $\eta^{MN} = (-+++++)$ 
for the metric and
\be \label{ee}
 \e^{01234} = - \e_{01234} =1, \quad \e^{012345} = - \e_{012345} =1
\ee
for the antisymmetric tensors.
The Hodge dual of a 3-form $G_{MNP}$ is defined by
\be \label{hodge}
 \tilde{G}_{MNP} := - \frac{1}{6} \e_{MNPQRS}\, G^{QRS}.
\ee  
Note  the minus sign in our definition of the Hodge dual follows from our 
convention
of the antisymmetric tensor \eq{ee} which says that the 6d orientation
is 
specified
by $dx^0 dx^1 \cdots dx^5$.
The abelian field strength is given by
\be
H_{MNP} = \del_M B_{NP} + \del_N B_{PM} + \del_P B_{MN} := \del_{[M} B_{NP ]}
\ee
and the self-duality equation reads 
\be \label{sd-a}
\Ht_{MNP} = H_{MNP}.
\ee

%c5 
In the Perry-Schwarz formulation, the self-dual tensor gauge field is represented
by a $5 \times 5$ antisymmetric tensor field $B_{\m\n}$.
%c2 do not appear. 
The  action reads
\be \label{S-PS-a}
%c7
S_{0}(B) = \frac{1}{2} \int d^6 x \, \left(
-\Ht^{\m\n} \Ht_{\m\n} + \Ht^{\m\n} \del_5 B_{\m\n}\right)
\ee
where
\be \label{Ht2}
\Ht^{\m\n} := \frac{1}{6} \e^{\m\n\r\l\s} H_{\r\l\s}, \qquad 
H^{\m\n\r} = -\frac{1}{2} \e^{\m\n\r\l\s} \Ht_{\l\s}.
\ee
The action has the second order  equation of motion
\be
\e^{\m\n\r\l\s} \del_\r (\Ht_{\l\s} -\del_5 B_{\l\s}) =0
\ee
which has the general solution
\be \label{soln-gen-a}
\Ht_{\l\s} -\del_5 B_{\l\s} =\Phi_{\l\s}
\ee
for some function $\Phi_{\l\s}$ such that $\del_{[\m} \Phi_{\l\s]}=0$.
It is easy to check that the action \eq{S-PS-a} 
is invariant 
\footnote{
This is under the usual assumption that fields, in this case $H_{\m\n\l}$, 
vanishes at infinity $|x^\m| = \infty$.
}
under the gauge symmetry
\be \label{PS-symm-a}
\d B_{\m\n} = \S_{\m\n}
\ee
for arbitrary $\S_{\m\n}$ such that $\del_{[\m} \S_{\n\l]} =0$,
or equivalently 
\be \label{PS-symm-b}
\d B_{\m\n} = \del_\m \vphi_\n -\del_\n \vphi_\m, \quad \mbox{for
arbitrary $\vphi_\m$}.
\ee 
%c4
This is the tensor gauge symmetry of the model.
An appropriate gauge fixing of this symmetry allows one to reduce the general
solution \eq{soln-gen-a} to the special form
\be \label{soln-spec-a}
\Ht_{\m\n} =\del_5 B_{\m\n}.
\ee
This is the  self-duality equation in this theory.

The action is manifestly 5d Lorentz invariant. Nevertheless
%c7 Perry and Schwarz were able to show that their 
the 
action is indeed invariant under 
an additional Lorentz transformation mixing the $\m$ directions 
with the 5 direction. The proposed  modified Lorentz transformation is
\be \label{mod-L-a} 
\d B_{\m\n} = (\L \cdot x) \Ht_{\m\n} - x_5 (\L \cdot \del)  B_{\m\n},
\ee
where  $\L_\m =\L_{5 \m}$ denote the corresponding infinitesimal transformation
parameters.
%c2 When put on shell using the equation of motion \eq{soln-spec-a}, 
% the modified Lorentz transformation is equal to 
% the standard Lorentz transformation.
%c1 \be \label{std-L-a}
% \d B_{\m\n} = ((\L \cdot x) \del_5 - x_5 (\L \cdot \del) ) B_{\m\n}
% + \L_\n B_{\m 5} - \L_\m B_{\n 5}
% \ee
% in the case $B_{\m 5} =0$. 
One can check that 
\be
[ \d_{\L_1}, \d_{\L_2}] B_{\m\n} = \d^{(5d)}_{\L_{\a\b}} B_{\m\n} + 
\del_\m \vphi_\n - \del_\n \vphi_\m 
\ee  
gives, apart from terms  that vanish on-shell \eq{soln-spec-a},
the expected 5d Lorentz transformation 
\be
\d^{(5d)}_{\L_{\a\b}} 
B_{\m\n} = \L_\m{}^\l B_{\l \n} -   \L_\n{}^\l B_{\l \m} 
+ x_\l \L^{\l\a} \del_\a B_{\m\n}  
\ee 
plus the gauge 
transformation \eq{PS-symm-b}.
The parameters are
\be
\L_{\m\n} = \L_{1 \m} \L_{2 \n} -  \L_{1 \n} \L_{2 \m}, \qquad 
\vphi_\n = x^\a \L_{\a \l} B_{\n}{}^\l.
\ee
Therefore the modified Lorentz 
transformation \eq{mod-L-a} does give rise to the desired 6d Lorentz group.

A couple of remarks follow %k1 follow 
concerning the Perry-Schwarz construction. 

\bit
 
\item[1.] We note that in the proof \cite{PS} 
of the invariance of the action \eq{S-PS-a}
under the Lorentz transformation \eq{mod-L-a}, various total derivatives terms 
in the variation of the action were dropped under the natural assumption that  
\be \label{asym-B}
\mbox{$\del_\l B_{\m\n} \to 0$ as 
$|x^M| \to \infty$ }.
\ee
%c3
Under the same assumption, 
the self-duality equation of motion \eq{soln-spec-a}
holds since  $H_{\m\n\l} \to 0$ at infinity.

\item[2.]
The Perry-Schwarz theory is based on the set of fields $B_{\m\n}$
which nevertheless is 6d Lorentz invariant. That it is possible 
to support the Lorentz symmetry without introducing the components
$B_{\m 5}$ is entirely due to the existence of the gauge symmetry
\eq{PS-symm-b} in the theory. In the manifestly Lorentz covariant 
formulation of PST \cite{pst}, 
the field $B_{\m\n}$ is extended
to $B_{MN}$. In addition an auxiliary scalar field $a$ is introduced with new gauge
symmetries that allow one to choose the gauge $B_{\m 5} =0$ and $a =x_5$.
In  this gauge, the  Perry-Schwarz action is obtained.

\item[3.] One may also combine the modified Lorentz transformation \eq{mod-L-a} 
with the gauge  transformation \eq{PS-symm-b} 
with a parameter $\vphi_\m = - x_5 B_{\m \k} \L^\k$ 
and obtain an equivalent form
of the modified Lorentz transformation
\be \label{mod-L-a2}
\d B_{\m\n} = (\L \cdot x) \Ht_{\m\n} - x_5 \L^\k  H_{\k \m\n},
\ee
which is written entirely in terms of the field strength. The check of
the invariance of the action under \eq{mod-L-a2} is included in the appendix.

\eit

\section{Action for Non-Abelian Self-Dual Two-Form on M5-Branes}

For simplicity, we will construct a theory of %the
% k7
the 2-form potential without scalars and fermions.
Supersymmetry is important and will be considered separately. 
For the gauge part,
motivated by the construction of \cite{chu}, 
we consider the addition of a set of 1-form gauge fields $A_M^a$ for a
gauge group $G$. 

\subsection{Non-Abelian action}

%c5 
Following 
%c7 
the above   discussion, 
we will give up manifest 6d Lorentz symmetry and
represent the self-dual tensor gauge field  by 
a $5\times 5$ antisymmetric field $B_{\m\n}$ in the 
adjoint.  Since 
there is no room for extra degrees of 
freedom in the (2,0)  tensor multiplets of M5-branes, therefore 
the gauge fields $A_M$  must be determined 
in terms of the tensor gauge fields.
It turns out we  need to take the Yang-Mills gauge field
to be a 5-dimensional field living in the 5d space $x^\m$, i.e.
$A_\m =A_\m(x^\l)$
%c6
\footnote{
We note that a 5-dimensional gauge field was also employed in 
\cite{ho}. However our construction differs from theirs 
in essential ways: a compactified
spacetime was considered in \cite{ho} and 
the gauge field was taken to be the zero mode of the tensor gauge field
$B_{\m 5}^{(0)}$. In our construction, we do not compactify the 
spacetime and $A_\m$ is given by an integrated expression \eq{FB} 
on shell.
We thank Pei-Ming Ho for a discussion on this point}.
Let us 
%c5 start with 
introduce the following non-abelian 
generalization of the Perry-Schwarz action
\be \label{S-PS-na}
S_0 = \frac{1}{2} \int d^6 x \, \tr \left(
-\Ht^{\m\n} \Ht_{\m\n} + \Ht^{\m\n} \del_5 B_{\m\n}\right),
\ee
where 
\be
H_{\m\n\l} = D_\m B_{\n\l} +  D_\n B_{\l\m} +D_\l B_{\m\n}
\ee
and
\be
\Ht^{\m\n} = \frac{1}{6} \e^{\m\n\r\l\s} H_{\r\l\s}
\ee 
is  the Hodge dual of $H_{\m\n\l}$. $H_{\m\n\l}$ obeys the modified Bianchi identity
\be
D_{[\m} H_{\n\l\r]} = \frac{3}{2} [ F_{[\m \n}, B_{\l\r]}].
\ee
The action $S_0$ is invariant under the 
Yang-Mills gauge symmetry
\bea
\d A_\m &=& \del_\m \L +[A_\m, \L], \quad \mbox{for arbitrary 
$\L =\L(x^\l)$}, \\
\d B_{\m\n} &=&  [B_{\m\n}, \L], \quad \d H_{\m\n\l} = [H_{\m\n\l}, \L] 
\eea
and the  following  ``tensor gauge symmetry''
%c9
\footnote{
Or equivalently
\be
\d_{T} B_{\m\n} = D_\m \L_\n- D_\n \L_\m 
\quad \mbox{for arbitrary $\L_\m(x^M)$ 
such that $[F_{[\m \n }\, , \, \L_{\l]}] =0$} . 
\ee
}:
\be
\d_{T} A_\m = 0, 
\ee
\be \label{gauge-symm-B}
\d_T B_{\m\n} = \S_{\m\n}, 
\qquad \mbox{for arbitrary $\S_{\m\n}(x^M)$ such
that $D_{[\l} \S_{\m\n]}=0$}.
\ee
It is $[\d_{T^{(1)}}, \d_{T^{(2)}}] =0$ and so
the tensor gauge symmetry is abelian.
Like the abelian case,
%c7 Perry-Schwarz, 
we will consider field configurations with vanishing
covariant derivatives at infinity:
%c3 
\be
D_\l B_{\m\n}\,, \; \del_5 B_{\m\n} \to 0 \quad \mbox{as} |x^M| \to \infty. 
\ee
It follows that $H_{\m\n\l}$ vanishes at infinity also.

An  important observation is that the condition for the vanishing of field 
strength at infinity:
\be
H_{\m\n\l} \to 0 , \quad \mbox{at $x_5 \to \pm \infty$}
\ee
is equivalent to the Bianchi identity of the gauge field $A_\m$ if
$F_{\m\n}$ is identified with the boundary value of $B_{\m\n}$,
e.g. $F_{\m\n} = B_{\m\n} ( x_5 =\infty)$. With the anticipation of the
self-duality equation of motion \eq{sd-na} in our theory, we will consider a
different constraint 
\be \label{FB}
F_{\m\n} = \int dx_5 \; \Ht_{\m\n}.
\ee
With the constraint \eq{FB}, there 
is no new degrees of freedom carried
by $A_\m$ 
\footnote{One may be tempted to use a Chern-Simons action to enforce
the gauge field to be auxiliary. However
unlike the 3-dimensional case where
a Chern-Simons gauge field is auxiliary and contains no local 
degrees of freedom, 
pure Chern-Simons gauge field in 5-dimension contains local degrees of 
freedom  \cite{HT1,HT2,HT3}. In the appendix, we review this argument
as well as the extension for Chern-Simons coupled to a conserved source.
}.
We will  implement \eq{FB} in the action by
introducing a 5-dimensional auxiliary field $E_{\m\n}(x^\m)$ and add the action 
\be
S_E = \int d^5 x \; \tr 
\left( (F_{\m\n} -\int dx_5 \; \Ht_{\m\n}  ) E^{\m\n}
\right).
\ee
The boundary condition of $E_{\m\n}$ will be taken as the trivial one
\be
E_{\m\n} \to 0 \quad\mbox{as} \quad |x^\l| \to \infty. 
\ee
$E_{\m\n}$ transforms 
under Yang-Mills and tensor gauge transformation 
as 
\be
\d E_{\m\n} = [E_{\m\n} ,\L] , \quad \d_T E_{\m\n} =0
\ee
and so $S_E$ is invariant. 
The action is also invariant under the gauge symmetry
\be \label{gauge-symm-E}
\d E_{\m\n} = \a_{\m\n}
\ee
for arbitrary $\a(x^\l) $ such that
\be
D_{[\m} \a_{\n\l]} =0, \quad D^\m \a_{\m\l} =0, \quad \mbox{and}
\quad \mbox{$\a \to 0$ as $|x^\l| \to \infty$}. 
\ee

All in all, we propose the following action for a non-abelian theory
of self-dual tensor
\be \label{S-tot}
S = S_0 +  S_E.
\ee
The action $S$ is Yang-Mills gauge invariant and tensor 
gauge invariant. 
It is also invariant under the gauge symmetry
\eq{gauge-symm-E} of $E_{\m\n}$.
Five dimensional Lorentz symmetry is manifest. 
We will show below 
this action leads to a  self-duality equation of motion.
We will also  demonstrate the existence of a
non-manifest 6d Lorentz symmetry in our theory and the connection to
5d Yang-Mills theory of multiple D4-branes through dimensional reduction on
a circle. The form of the constraint \eq{FB} is inspired by the
analysis of this reduction.

\subsection{Properties}

\subsubsection{Self-duality}

The equation of motion of $E_{\m\n}$ gives the constraint
\be \label{FB1}
F_{\m\n} = \int dx_5 \; \Ht_{\m\n}.
\ee
This has to satisfy the Bianchi identity 
\be \label{Bid}
\e^{\m\n\r\l\s}D_\r F_{\l\s} =0.
\ee 

For $B_{\m\n}$, we have
\be \label{dS0}
\d S_0 = \frac{1}{2} \int\e^{\m\n\r \l\s} \d B_{\m\n} 
D_\r (H_{\l\s} -\del_5 B_{\l\s})
\ee
and hence the equation of motion
\be \label{eom-B}
\e^{\m\n\r\l\s} D_\r (\Ht_{\l \s} - \del_5 B_{\l\s} + E_{\l\s}
) =0,
\ee
Integrating it over $x_5$, we get
\be \label{DE1}
D_{[\r} E_{\l\s]} =0.
\ee 
In fact
$
\int dx_5 \, \e^{\m\n\r\l\s}D_\r (\Ht_{\l \s} - \del_5 B_{\l\s}) = 0
$
where we have used \eq{FB1} and the Bianchi identity of $F_{\m\n}$,
and we have assumed that $H_{\m\n\l}$ vanishes at $x^5 = \pm
\infty$. Our claim follows from the fact that 
$E_{\l\s}$ is independent of $x_5$.
%c4
As a result, the equation \eq{eom-B} 
%c8
reads
\be \label{eom-B1}
\e^{\m\n\r\l\s} D_\r (\Ht_{\l \s} - \del_5 B_{\l\s}) =0
\ee
and has the general solution 
\be \label{gen-soln-na}
\Ht_{\l \s} - \del_5 B_{\l\s} = \Phi_{\l\s},
\ee
where
\be
D_{[\l} \Phi_{\m\n]} =0.
\ee 
%c4
Therefore with an appropriate fixing of the gauge symmetry \eq{gauge-symm-B},
one can always reduce the second order
equation \eq{gen-soln-na} to the first order form
\be \label{sd-na}
\Ht_{\m\n} =\del_5 B_{\m\n}. 
\ee
This is the form of the self-duality equation in our theory.

The equation \eq{sd-na} 
implies that on-shell, $F_{\m\n}$ is simply given in terms of  
the boundary values of $B_{\m\n}$:
\be
F_{\m\n} = B_{\m\n}(x_5 =\infty) -  B_{\m\n}(x_5 =-\infty),
\ee
and Bianchi identity is satisfied since the field strength vanishes at
infinity.
Finally, the equation of motion for $A_\m$ gives
%c8
\be \label{DE2}
D^\m E_{\m\n} - 
\frac{1}{4} \int dx_5 \; \e_{\n}{}^{\a\b\g\d} [B_{\a\b}, E_{\g\d}]
= -\frac{1}{2} 
\int dx_5 \; \e_{\n}{}^{\a\b\g\d} [B_{\a\b}, \del_5 B_{\g\d} 
- \frac{1}{2} \Ht_{\g\d}]:=J^\n.
\ee
We note that
as a result of the self-duality equation of motion
\eq{sd-na}, the ``current'' is covariantly conserved
$D_\l J^\l =0$ . Of course  \eq{DE2} is consistent with this. 
 
Summarizing, the equations of motion in our theory are the auxiliary
equation for $A_\m$ \eq{FB},  the self-duality equation \eq{sd-na} and
the equations \eq{DE1} and \eq{DE2} for $E_{\m\n}$. Note that on
eliminating $A_\m$ using \eq{FB}, the self-duality 
equation \eq{sd-na} is self-interacting and is completely
independent of $E_{\m\n}$. 

The counting of the degrees of freedom in our theory goes as
follows. The equation of motion \eq{FB1} says $A_\m$ is auxiliary and
is determined entirely in terms of $\Ht_{\m\n}$. Using this, the
action $S$ can be written as a nonlocal action 
in terms of expansion in powers of $B_{\m\n}$.  
At the quadratic level, the action is simply
given by ${\rm dim} G$ copies of the Perry-Schwarz action, plus the
action $S_E$. For small
field strengths, we can take the higher order terms as small corrections
and we can count the degrees of freedom using the linearized
theory. In this limit, $A_\m =0$ and 
the tensor gauge symmetry and the  self-duality equation of motion
are precisely those of the original Perry-Schwarz theory. Thus we
obtain  $3 \times {\rm dim} G$ degrees of freedom in $B_{\m\n}$. 
As for $E_{\m\n}$, the linearized equations of motion are
\be
\del_{[\m} E_{\n\l]} =0, \quad \del^\m E_{\m \n} =0,
\ee
and there is the gauge symmetry \eq{gauge-symm-E} with the parameters $\a_{\m\n}$
satisfying, in this case, 
\be
\del_{[\m} \a_{\n\l]} =0, \quad \del^\m \a_{\m \n} =0.
\ee
Since $E_{\m\n}$ and $\a_{\m\n}$ also satisfy the same (vanishing)
boundary  condition at infinity, so we can use the gauge symmetry to remove
the $E_{\m\n}$ field completely. 
This is compatible with the fact $E_{\m\n}$
was introduced as an auxiliary field to implement the constraint
\eq{FB}. All in all, our theory contains  $3 \times {\rm dim} G$ degrees of 
freedom as required by (2,0) supersymmetry

We remark that when $B_{\m\n}$ is diagonal with distinct diagonal elements such
that the gauge group is broken down to $U(1)^r$ ($r$ 
is the rank of the gauge group), 
our action reduces to a sum of $r$ copies of the  abelian
Perry-Schwarz theory and describes 
%c10
the gauge sector of $r$ separated M5-branes. 
More generally,
%c10
once the scalar and fermion fields are included in the theory,
one can have a system of lumps of coincident M5-branes, BPS or 
non-BPS relative to each other; and 
as usual, the 
pattern of symmetry breaking as well as the interacting dynamics of M5-branes 
can be studied.

\subsubsection{Lorentz symmetry}

Our action is manifestly 5d Lorentz invariant. 
It is straightforward to check that it is not invariant under the
modified Lorentz transformation 
%c4
\eq{mod-L-a} or \eq{mod-L-a2}.
See appendix A for the check. 
Let us proceed by further modifying 
the Lorentz transformation.  We observe that 
the equation \eq{dS0} for the variation of $S_0$ 
under  a general variation of $\d B_{\m\n}$ can be rewritten as
\bea \label{dS}
\d S_0 &=& 
%c7 typo
%c7 \frac{1}{2} 
\int d^6 x \; \tr \left[
\Delta B^{\m\n} \Ht_{\m\n} \right],
\eea
where
\be
\Delta B^{\m\n} := \del_5( \d B^{\m\n}) - \frac{1}{2} \e^{\m\n\a\b\g} 
D_\a (\d B_{\b\g}).
\ee
It is interesting to note that
\be \label{DB0}
\Delta B_{\m\n} = - \d (\Ht_{\m\n} - \del_5 B_{\m\n}), 
\ee
which is just the variation of the self-duality equation of motion. 
%c8  Therefore
% the self-duality equation is invariant up to a symmetry of the action.

Taking $\d B_{\m\n}$ now as the 5-$\mu$ Lorentz transformation, it is clear 
that
the action will be invariant if the variation satisfies 
%c7 the equation
%c7 \be \label{DB}
$ \Delta B_{\m\n} =0$. 
This is a sufficient condition,
%c8
but not necessary. In fact $\D B_{\m\n} \neq 0$
for the abelian case \eq{mod-L-a2},  
nevertheless $S_0$ is invariant. So let us consider
a general transformation of the form
\be \label{mod-L-phi}
\d B_{\m\n} = (\L \cdot x) \Ht_{\m\n} - \l x_5 \L^\k H_{\k \m\n} +
\L^\k \phi_{\m\n\k} := \d_{(1)} B_{\m\n} +  \d_{(2)} B_{\m\n},
\ee
where $\l$ is a constant and 
$\phi_{\m\n\k} = - \phi_{\n\m \k}$ is a quantity to be determined by demanding
$S_0$ to be invariant. 
We have denoted the first two variation terms 
by $\d_{(1)} B_{\m\n}$  and the third term by $\d_{(2)} B_{\m\n}$.
By redefining $\phi_{\m\n\k}$ with an appropriate shift, 
one can bring $\l$ to any value one wants. This freedom will turn out to be 
convenient.

The variation of $S_0$ under $\d_{(1)} B_{\m\n}$ is
\be\label{dS1}
\d_{(1)} S_0 = \int \left[
\frac{\l}{2} x_5 \e^{\m\n\a\b\g} D_\a H_{\b\g \k}\L^\k 
+\frac{\l -1}{4} \L_\r \Ht_{\a\b}  \e^{\r \a\b\m\n} \right] \Ht_{\m\n}.
% k4 typo of sign
\ee
For $\l =1$, the result in the appendix is recovered. For the moment, let us keep 
$\l$ arbitrary. Since \eq{dS1} is of the form of \eq{dS}, therefore it can be 
cancelled with $\d_{(2)} B_{\m\n}$ if $\phi_{\m\n\k}$ satisfies 
\be\label{DE-phi}
\del_5 \phi_{\m\n\k} - \frac{1}{2}\e_{\m\n}{}^{\a\b\g} D_\a \phi_{\b\g \k} =
-\frac{\l}{2} x_5 \e^{\m\n\a\b\g} D_\a H_{\b\g \k} %\L^\k 
-\frac{\l -1}{4}  \Ht^{\a\b}  \e_{\k \a\b\m\n}:=J_{\m\n\k}.
% k4 typo of sign
% k4 indices balanced
% k4 gauge parameter on RHS removed.
\ee
In addition, we impose the boundary condition 
\be\label{BC-phi}
\mbox{$\phi_{\m\n\k}$ vanishes as $|x_5| \to \infty$}.
\ee 
A solution can always be written down using the Green function 
technique for general $J_{\m\n\k}$. 
Let $G^{ab}_{\m\n,\m'\n'}(x,y)$ be the Green function which satisfies
\be
\del_5 G^{ab}{}_{\m\n}^{\m'\n'}
- \frac{1}{2}\e_{\m\n}{}^{\a\b\g} (D_\a^{(y)})^a{}_c G^{cb}{}_{\b\g}^{\m'\n'}
% k4 typo of lower indices of G  
= \d^{\m'\n'}_{\m\n} \d^{ab} \d^{(6)}(x-y)
\ee
and the boundary condition
\be
G^{ab}{}_{\m\n}^{\m'\n'}(x,y) =0, \quad |x_5| \to \infty.
\ee
Here $x = (x^M)$ and $(D_\a)^a{}_c = \del_\a \d^a{}_c + (\At_\a)^a{}_c$ where
$(\At_\a)^{ac} := f^{abc} A_\a^b$. Then  
\be \label{phi-Green}
\phi^a_{\m\n\k} = \int dy\; G^{ab}{}_{\m\n}^{\m'\n'}(x,y) J_{\m'\n' \k}^b(y)
\ee
satisfies both \eq{DE-phi} and \eq{BC-phi}. As a result, if also
\be \label{L-A}
\d A_\m =0, 
\ee
then $S_0$ is invariant.
So far this works for any $\l$. 

Next let us examine the action $S_E$. It follows from \eq{mod-L-phi} that
\be \label{dHt}
\d \Ht_{\m\n} = \del_5 \phi_{\m\n\k} \L^\k 
% k5 \La^\k missing  
+ 
\frac{\L\cdot x }{2} \e_{\m\n}{}^{\a\b\g} D_\a \Ht_{\b\g}
+ \frac{\l+1}{4} \e_{\m\n}{}^{\a\b\g}\L_\a \Ht_{\b\g},
\ee
where we have used the differential equation \eq{DE-phi}. Therefore $S_E$ is 
invariant if we take $\l =-1$ and if $E_{\m\n}$ transforms as
\be \label{L-E}
\d E_{\m\n} = \frac{1}{2} \e_{\m\n}{}^{\a\b\g} D_\a(( \L\cdot x) E_{\b\g}).
% k4 indices balanced.
\ee
All in all, our action is invariant under the transformation
\eq{mod-L-phi}, \eq{L-A} and \eq{L-E}.

In general the 
%c9 
Lorentz invariance of the action implies that the equations of motion 
(i.e. \eq{FB}, \eq{eom-B1} \eq{DE1} and \eq{DE2})
are automatically 
%c9 
Lorentz invariant, up to terms vanishes on shell and terms that 
can be interpreted as any other symmetry transformations of the theory. However
since the self-duality equation \eq{sd-na} is obtained by a gauge fixing, it
is not  guaranteed to be Lorentz invariant.  
In fact, the transformation \eq{mod-L-phi} implies that 
\be \label{dHB}
\d(\Ht_{\m\n} -\del_5 B_{\m\n} ) = \frac{\L\cdot x}{2} 
\e_{\m\n}{}^{\a\b\g} D_\a \Ht_{\b\g} - (\L \cdot x) \del_5 \Ht_{\m\n} 
-\del_5(x_5 H_{\m\n\k}\L^\k).
\ee
This gives in \eq{dS} $\d S_0 =0$ as expected.
Using the self-duality equation \eq{sd-na}, the first and second term of 
\eq{dHB} actually cancel and so
\be
\d(\Ht_{\m\n} -\del_5 B_{\m\n} ) = -\del_5(x_5 H_{\m\n\k}\L^\k) 
+ \mbox{EOM},
\ee
where EOM denotes terms vanish when the equation of motion \eq{sd-na} is used. 
One can rewrite this further by using the equation of motion
and obtains
\be \label{vHB}
\d(\Ht_{\m\n} -\del_5 B_{\m\n} ) = 
\frac{1}{2}\e_{\m\n\k}{}^{\a\b} \L^\k (\Ht_{\a\b} + 2 x_5 \del_5 \Ht_{\a\b}) + 
x_5 \L^\k D_\k \Ht_{\m\n} + D_{[\m} \vphi_{\n ]}
+ \mbox{EOM},
\ee
where $\vphi_\n = x_5 \Ht_{\n\k} \L^\k$.
Now the first and second term on the RHS of  \eq{vHB} respectively gives zero when 
substituted into \eq{dS} and so they corresponds to symmetry transformations
of the action $S_0$ 
\footnote{More specifically, the symmetry transformations are given by 
$\d B_{\m\n} = \phi_{\m\n\k}\L^\k$ where $\phi_{\m\n\k}$ is given by 
\eq{phi-Green} with 
$J_{\m\n\k}$ specified by the first and second term of the RHS of \eq{vHB}
respectively. 
}.
For the abelian case, the third term corresponds to the symmetry transformation 
$\d B_{\m\n} = \del_{[\m} \a_{\n]}$ of $B_{\m\n}$ and since $S_E$ decouples 
from the theory, so we obtain that the self-duality equation is Lorentz invariant
up to terms vanishes on shell and terms that correspond to a symmetry 
transformation of the theory. However the above analysis breaks down in the 
non-abelian case and so we conclude that the self-duality equation of motion is
not Lorentz invariant.  
We emphasize that the loss of Lorentz invariance in \eq{sd-na} is simply because
it is a gauge fixed equation of motion. This is not surprising. For example, 
Yang-Mills equation of motion in the Coulomb gauge is not Lorentz invariant.
The use of the self-duality equation is important for obtaining the correct 
counting on the degrees of freedom in the theory. However the use of the 
ungauge-fixed version \eq{eom-B1} may be useful for some other purposes, 
for example, supersymmetry.

%c7
If we compute the algebra of
commutator $[\d(\L_\m^{(1)}), \d(\L_\m^{(2)})]$ for the physical field $B_{\m\n}$, 
we get the standard 5d Lorentz
transformation plus an additional transformation. This additional
transformation is quite complicated but is a symmetry of the action since we know
already the action is invariant under the 5d Lorentz transformation
and is invariant under $[\d(\L_\m^{(1)}), \d(\L_{\m}^{(2)})]$.
Therefore we can interpret \eq{mod-L-phi} as a modified Lorentz symmetry.
Note that the form of the transformation laws \eq{L-A} and \eq{L-E} are 
quite non-standard
but they are compatible with the auxiliary 
% k3 typo auxiliary
nature of these fields.

We note that as $\phi_{\m\n\k}$ is determined explicitly 
% k3 typo explicitly
as an integrated expression over the Green function, 
the transformation \eq{mod-L-phi} is non-local in the fields.
It is now clear that the different choices of $\l$ simply 
correspond to different non-local form of the transformation \eq{mod-L-phi}. What 
we have shown is that one can make the action invariant by using a transformation
law that has a nonlocal piece that is based on a local part with 
the particular choice of $\l =-1$. For the
abelian case, we know the Lorentz transformation \eq{mod-L-a2} 
%c9 
is locally represented in terms of $A_\m$ and $B_{\m\n}$;
and corresponds to 
$\l=1$ and $\phi_{\m\n\k}=0$. Let us demonstrate that this 
is equivalent to having $\l =-1$ and 
a nontrivial $\phi_{\m\n\k}$ as determined above. 
To see this, the equation 
\eq{DE-phi} reduces in the abelian case to
\be
\del_5 \phi_{\m\n\k} - \frac{1}{2}\e_{\m\n}{}^{\a\b\g} \del_\a \phi_{\b\g \k} =
% k4 indices balanced.
x_5 \del_\k \Ht_{\m\n} - H_{\m\n\k}.
\ee
Let us put 
$\phi_{\m\n\k} = -2x_5 H_{\m\n\k} + \vphi_{\m\n\k}$
and so
\be
\del_5 \vphi_{\m\n\k} - \frac{1}{2}\e_{\m\n}{}^{\a\b\g} \del_\a \vphi_{\b\g \k} = 
% k4 typo 1/6 -> -1/2 ?
-\frac{1}{2}  \e_{\m\n\k}{}^{\a\b}(\Ht_{\a\b}  + 2 x_5 \del_5 \Ht_{\a\b})
- x_5 \del_k \Ht_{\m\n}.
\ee
Now the right hand side of this equation when substituted into \eq{dS} 
actually leaves $S_0$ invariant. Therefore as explained above, 
$\vphi_{\m\n\k}$  represents a symmetry and we
recover \eq{mod-L-a2} up to a symmetry transformation.

The Lorentz symmetry we proposed is nonlocal and 
is quite different from the usual representation of 
a symmetry in terms of local fields, but it seems  this is what 
is needed for 
multiple M5-branes
\footnote{We thank Pei-Ming Ho and Yutaka Matsuo for emphasizing 
% k3 typo emphasizing
the 
nonlocal nature of our proposed Lorentz transformation and
for a discussion on this point.}. 
In fact, 
nonlocal symmetry is not uncommon in string theory. For example, the 
spacetime Lorentz symmetry in the light cone gauge string theory is nonlocal in the 
worldsheet coordinate \cite{GSW}.  There the nonlocality arises since a 
Lorentz 
transformation will generally bring one out of the lightcone gauge and so a
worldsheet reparametrization (turns out to be nonlocal) 
is needed in order to restore the gauge condition.
For us,  we are in a formulation without the $B_{5\m}$ fields. Since 
a standard 5-$\mu$ Lorentz transformation will turn $B_{\m\n}$ to $B_{5\m}$,  
we suspect that the reason of having a modified Lorentz 
symmetry is similarly due to a compensating gauge transformation in 
a covariant formulation.
%c10
In the abelian (free) 
case, the modification is not so drastic and the modified Lorentz 
transformation is still local. But this is not the case for the non-abelian case
as we found here.
To check our suspicion,  
it is needed to construct the covariantized theory.
It is remarkable that for the abelian 
case, PST \cite{pst} 
were able to provide a Lorentz covariant formulation by introducing additional 
auxiliary fields (scalar field $a$ and the $B_{5\m}$ components).
It will be 
very interesting to
covariantize our construction by following a similar  construction of 
PST and it is possible that  
the employment of additional auxiliary fields would allow for 
a local representation of the Lorentz symmetry.

\subsubsection{Reduction to D4-Branes}

Let us consider a compactification of $x_5$ on a circle of radius $R$. 
The dimensional reduced action reads
\be
S = \frac{2 \pi R}{2} \int d^5 x \, \tr 
\left( 
-\Ht_{\m\n}^2 +(F_{\m\n} - 2 \pi R \Ht_{\m\n}) E^{\m\n}  
\right) 
\ee
This form of action has been considered in \cite{chu} as a dual
formulation of 5-dimensional Yang-Mills theory. In fact, if we
integrate out $E_{\m\n}$, we obtain
%c4
the expected relation
\be \label{w0}
F_{\m\n} = 2 \pi R \Ht_{\m\n}.
\ee
Eliminate $\Ht_{\m\n}$ using the constraint, we obtain the standard
5d Yang-Mills action
\be
S_{YM}= -\frac{1}{4 \pi R} \int d^5 x \; \tr \; F_{\m\n}^2 . 
\ee
%c7 
This is however not the complete answer. In fact if we look at the path integral
and integrate out $E$ first, we obtain
\be
\int [DA] [DB] [DE] e^{-S} = \int [DA] [DB] e^{-S_{YM}} 
\d( F_{\m\n}- 2\pi R \Ht_{\m\n}) = \int [DA] e^{-S_{YM} - S'}, 
\ee
where $S'=S'(A)$ is a measure contribution obtained
from integrating out the delta functional constraint and 
then rewritten in terms of $A_\m$. The direct 
determination of $S'$ is nontrivial but it has to satisfy a consistency 
condition: the condition
\be
D_\m F^{\m\n} = - \frac{\pi R}{2} \e^{\n\a\b\g\d} [F_{\a\b}, B_{\g\d}] 
\ee
which follows from \eq{w0} should be obtained as an equation of motion in the 
5d theory. As a result, $S'$ has to satisfy
\be
\frac{\d S'}{\d A_\n} =  \frac{1}{2} \e^{\n\a\b\g\d} [F_{\a\b}, B_{\g\d}]
\ee
with $B_{\m\n}$ understood to be a function of $A_\m$ obtained by solving
the duality relation \eq{w0}.

The 5d theory is thus given by the action $S_{5d} = S_{YM} + S'$.
The action $S_{YM}$ corresponds to the expected form of the  Yang-Mills coupling
\be
g_{YM}^2 = R
\ee
and the gauge group in our construction  is to be
\be
G = U(N)
\ee
for a system of $N$ M5-branes.  The reproduction of the 5d Yang-Mills action
gives further support that  our construction gives a description 
%c10
of the gauge sector of a system of 
multiple M5-branes. 
The action $S'$ describes a correction term to the Yang-Mills theory
which appears to be of high derivative in nature since $[F,B] \sim D D B$ and 
$B$ is of the order of $F$ from \eq{w0}). 
In the abelian case, Perry and 
Schwarz has also constructed the nonlinear 
%c10 M5-brane 
five-brane action that gives the
%c10 
$U(1)$ DBI action of D4-brane upon dimensional reduction.   
It would be interesting to work out $S'$ in more details and 
see whether it captures the non-abelian DBI action
\cite{dbi} in some way.

We remark that the 
necessity of non-locality in the M5-branes action has also been 
argued by Witten \cite{wit}. 
He observed that conformal invariance of the M5-branes
theory implies that upon double dimensional reduction to five dimensions, the
5 dimensional action should be proportional to 
\be \label{w1}
\frac{1}{R} \int d^5 x.
\ee
On the other hand, one should get
\be \label{w2}
\int d^6 x = 2\pi R \int d^5 x
\ee
as a result of integrating over the $x_5$ direction for 
a standard reduction of a local action, In our analysis above, we see that 
both $R$-dependence are correct and the trick to arrive from 
\eq{w1} to \eq{w2} is due to the simple $R$ dependence in 
the constraint \eq{w0}.

In principle one could consider compactification in the other spacelike
directions and one should get the same 5d YM action. However this is already 
non-trivial for the Perry-Schwarz action \cite{PS}  (or the 
Henneaux-Teitelboim action \cite{HT})  and implies
the existence of a 
symmetry of the D4-branes action which involves a non-local
field redefinition. For a single M5-brane, this symmetry 
can be made explicit in a covariant  
PST-like formulation in which both, the vector field $A_\m$ and
the two-form field $B_{\m\n}$ are present and related to each other, on the 
mass-shell, by the duality condition which follows from the  action. 
See for example \cite{dual} for the case of the duality-symmetric 
formulation of $D=11$ supergravity 
with $A_3$ and $A_6$ gauge fields. The construction is completely generic and
can be extended immediately to arbitrary $D$ dimensional spacetime any pair
of duality related fields of rank $p$ and $(D-p-2)$ whose field strengths
are dual to each other on the mass shell
\footnote{We thank Dmitri Sorokin for explaining this to us.}.
It would be interesting to extend this construction to the non-abelian case.

\section{Discussions}

In this paper, we have constructed a theory of non-abelian tensor fields 
with the properties 
that: 
\bit 
\item[1.] the action admits a self-duality equation of motion, 
\item[2.] the action has 
manifest 5d Lorentz symmetry and a modified 6d Lorentz symmetry, 
\item[3.] on dimensional
reduction, the action gives 
%c7 precisely 
the 5d Yang-Mills action plus certain higher derivative corrections. 
\eit
Based on these properties, we propose our action to 
be the bosonic theory describing the 
gauge sector of coincident M5-branes in flat space. A special
feature of our construction is that the tensor gauge symmetry is
abelian although the theory is still fully interacting. This is an
interesting difference between the self-interaction of 
Yang-Mills gauge fields  
and the self-interaction of 2-form gauge fields in our construction.
%c10
It remains to be seen whether this is still the case in the Lorentz covariant
formulation of the theory. 

%c3
We note that conformal symmetry 
rules out the possibility of 
a Yang-Mills action, but a 5d Chern-Simons action is allowed
for the gauge field $A_\m$:
\bea \label{S-CS}
S_{CS} = \frac{k}{24 \pi^2}  \int d^5 x \; \e^{\m_1 \cdots \m_5}  
\tr \left(
A_{\m_1} \del_{\m_2} A_{\m_3} \del_{\m_4} A_{\m_5} 
+ \frac{3}{2} A_{\m_1} A_{\m_2} A_{\m_3} \del_{\m_4} A_{\m_5}
\right. \nn\\
\left. + \frac{3}{5} A_{\m_1} A_{\m_2} A_{\m_3} A_{\m_4} A_{\m_5} 
\right).
\eea
%c3 remove statements about counting for CS 
The inclusion of the Chern-Simons action seems to corresponds to a kind of  
M-theory compactification as 5d Chern-Simons term  naturally arises and plays 
a very important role 
in certain kinds of M-theory compactification on  
Calabi-Yau manifolds, see for example 
\cite{5dcs1}, \cite{5dcs2}. In this case, the level 
$k$ may 
%c5 be 
corresponds  to a parameter describing a kind
of fibered Calabi-Yau compactification. It will certainly be helpful to 
have the full supersymmetric theory from which one may obtain
the moduli space interpretation from the scalar sector \cite{chub}.

Our construction is in principle only a low energy effective 
description for a system of 
coincident M5-branes. If one is lucky, the (2,0) supersymmetric 
completion may give a well-defined
quantum theory as in the case of BLG \cite{BLG} and ABJM theories \cite{ABJM} 
for multiple M2-branes and the $\cN=4$ SYM theory for multiple D3-branes.  
This is another strong reason to construct the supersymmetric
completion.

%c10
To construct the supersymmetric theory, one needs to include scalar
% k6 scalar
 fields and 
fermions in the adjoint of $U(N)$. For (2,0) supersymmetry, all these fields 
are sitting in the tensor multiplet. 
Since there is no Yang-Mills multiplet in (2,0) supersymmetry,
the Yang-Mills gauge field must be a supersymmetric singlet.
This is rather difficult to implement. On the other hand, it is possible that
only a fraction of the (2,0) supersymmetry, i.e. (1,0) supersymmetry, is
visible in the classical action 
%c10 
of multiple M5-branes, 
and full supersymmetry can be seen only
nonperturbatively as in the ABJM theory \cite{ABJM}. 
With respect to (1,0) supersymmetry, the (2,0)
tensor multiplet is simply the sum of a (1,0) tensor multiplet 
and a (1,0) hyper-multiplet. Moreover, one should employ a (1,0) Yang-Mills 
multiplet as an auxiliary multiplet. 
%c10 
The recent results of (1,0) 
superconformal theories \cite{sezgin} should be useful in this regard. 

%c10
However even before one enters into the details, a simple observation
already indicates that 
the supersymmetric theory is going to be highly nontrivial.
In six dimensions, scalar field has dimension 2. Conformal invariance plus locality 
imply  
% k6 imply 
that the potential term $V$ for the scalar fields has to be cubic. 
However a nonvanishing 
cubic potential has no ground state and this is not compatible with 
supersymmetry
\footnote{
This observation is also shared independently by 
David Berman, Neil Lambert, David Tong. 
}. 
This means the potential term, if nonvanishing, will need 
to be nonlocal. 
For example, potential of the schematic form 
$V \sim \phi^4/ |\phi|$ or $V \sim \int dx_5 \int dx_5 \; \phi^4$
could avoid the problem of not having a ground state. It is amusing
that the later 
form of the potential has a close resemblance 
with the scalar interaction term in
\cite{lam1} 
\footnote{We thank Neil Lambert for pointing out this resemblance.
} 
if one exchanges $C_\mu \sim \d^5_\m \int dx_5$, both of which are
of dimension -1.

It 
%c9 is also
would be  
interesting to understand the 
connection between our description 
and the proposed SYM description of M5-branes 
\cite{dou,lam2}. In particular an 
understanding of how a non-abelian 2-form gauge field would 
arise in the Yang-Mills description is needed. Incidentally, based on a
fluctuation analysis of D1-branes around a large RR 3-form flux background, 
a matrix model description for M5-branes in a background $C$-field was 
suggested in \cite{CG} and there is the same question 
of how to extract a $B$-field from the matrix variables. This problem
may be compared with the problem of extracting the spacetime fields
%c10
and their dynamics, 
particularly the gravity field,  
from the matrix model \cite{BFSS,IKKT}. 
See for example \cite{t1,t2,t3}. 
Lessons drawn from those analysis may be useful here.

Our theory is based on fields in the adjoint of $U(N)$, 
i.e. taking $N^2$ values. Naively this is different from the
$N^3$ counting from entropy argument \cite{m5-S}. To understand the counting, 
it will be important to understand the dynamics of the theory properly.
See for example \cite{lee} for some recent interesting analysis performed 
on the 5d SYM theory and a class of 6d SCFT in the Coulomb phase. 

\appendix

\section{Counting of degrees of freedom 
in the Perry-Schwarz theory}

We give a pedagogical and explicit counting of the degrees
of freedom in the Perry-Schwarz theory. 
The Perry-Schwarz theory initially has the equation of motion
\be
\e^{\m\n\r\l\s} D_\r (\Ht_{\l\s} -\del_5 B_{\l\s}) =0
\ee
Using the gauge symmetry
\be
\d B_{\m\n} = \del_\m \L_\n - \del_\n \L_\m,
\ee
one can fix the equation of motion to the linear form
\be \label{Hmn} 
\Ht_{\m\n} = \del_5 B_{\m\n}.
\ee
Doing so we are left with a $x^5$-independent residual symmetry. Now
$\del^\m B_{\m\n}$ is $x_5$ independent as a result of \eq{Hmn}. Using
the residual symmetry, one can fix it to be zero
\be \label{gc1}
\del^\m B_{\m\n} =0.
\ee 
Differentiating \eq{Hmn} with respect to $x_5$ and use \eq{gc1}, we
obtain that $B_{\m\n}$ is massless as expected, $\Box B_{\m\n}
=0$. Now \eq{gc1} gives 4 independent conditions on the
10 components of $B_{\m\n}$. Using the self-duality condition, 
we have in total $(10-4)/2 =3$
degrees of freedom.

\section{Variation of $S_0$ under Lorentz transformation}

In this appendix, we show that the
non-abelian Perry-Schwarz action
\be
S_0 = \frac{1}{2} \int d^6 x \, \tr \left(
-\Ht^{\m\n} \Ht_{\m\n} + \Ht^{\m\n} \del_5 B_{\m\n}\right),
\ee
is not invariant
under the straight-forward non-abelian generalization of the 
Lorentz transformation \eq{mod-L-a2} (i.e. with $\phi_{\m\n\k} =0$ 
in \eq{mod-L-phi}):
\bea \label{L0}
\d B_{\m\n} &=& (\L \cdot x) \Ht_{\m\n} - x_5 \L^\k H_{\k \m\n}, \\
\d A_\m & =& 0.
\eea
It is
\be
2 \d S_0 = \int  \e^{\m\n\r\l\s} \tr \; 
\Big[ 
\big(
\underbrace{(\L \cdot x) \Ht_{\m\n}}_{1} - 
\underbrace{x_5 \L^\k H_{\k \m\n}}_{2} 
\big)
\big(
\underbrace{D_\r \Ht_{\l\s}}_{a} - \underbrace{D_\r \del_5 B_{\l\s}}_{b} 
\big)
\Big].
\ee
The contributions are, respectively,
\bea
\mbox{(1a)} &=& -\frac{1}{2} \int   \tr\;  (\e^{\m\n\r\l\s}
\L_\r \Ht_{\m\n} \Ht_{\a\b}) + \mbox{tot.} \; , \\
\mbox{(2b)} &=&  - \int \tr\;  
(\e^{\m\n\l\a\b}  x_5 \Ht_{\a\b} \del_5 \Ht_{\m\n} \L_\l) =  
\frac{1}{2} \int  \tr\; (\e^{\m\n\r\l\s} 
\L_\r \Ht_{\m\n} \Ht_{\a\b}) + \mbox{tot.} \; ,\\
\mbox{(1b)} &=&   - 2 \int (\L\cdot x)  \tr  (\Ht_{\m\n} \del_5 \Ht^{\m\n}) 
=   \mbox{tot.} \; ,\\
\mbox{(2a)} &=& \int 2 x_5 \L^\k \, \tr\; ( H_{\k\m\n} D_\r H^{\m\n\r})
= \int 2x_5 \L^\k\, \tr \;( \frac{1}{3} H^{\r\m\n} D_{[\k} H_{\r\m\n]})
+ \mbox{tot.} \;,
\eea
where tot. stands for total derivative terms and we have used
\be
D_{[\k} H_{\r\m\n]} = D_\k H_{\r\m\n} - D_{[\r} H_{\m\n] \k}
\ee
in simplifying (2a).
We see that (1a) cancels (2b). In the abelian case, the term (2a) is 
zero due to the vanishing Bianchi identity $\del_{[\k} H_{\r\m\n]} =0$. This is
not so  for the non-abelian case and so $S_0$ is not invariant under
\eq{L0}. 
%c4
It is straightforward to see that $S_0$ is also not invariant under
\be
\d B_{\m\n} = (\L \cdot x) \Ht_{\m\n} -x_5 (\L \cdot D) B_{\m\n}.  
\ee

\section{Counting of degrees of freedom for 
Chern-Simons theory}

We will start with a review of the counting of degrees of freedom for pure 
Chern-Simons theory performed in \cite{HT1,HT2}. 
Then we extend the analysis to the case where the 
Chern-Simons theory is coupled to a covariantly conserved current.
The details of the counting is not important for our results. They are included
here for completeness.

\subsection{Pure Non-Abelian Chern-Simons theory}

Consider the five dimensional (dimension $D = 2n+1$, $n=2$ here)
Chern-Simons action 
\be
	S_{\text{CS}} = \int_M \cL_{\text{CS}},\qquad 
	\text{with}\quad d\cL_{\text{CS}} = g_{abc}F^a\wedge F^b \wedge F^c
\ee
where $g_{abc}$ is the symmetric invariant tensor of the gauge group and 
$a=1,\cdots, \cN$ with $\cN$ being the dimension of the gauge group.
The equation of motion 
\be
g_{aa_1a_2} F^{a_1}_{\m_1\m_2}F^{a_2}_{\m_3\m_4} \e^{\m_1\m_2\m_3\m_4\l} = 0
\ee
can be decomposed into 
\be
\begin{cases}
k_a\equiv g_{aa_1a_2} F^{a_1}_{i_1i_2}F^{a_2}_{i_3i_4} \e^{i_1i_2i_3i_4} = 0,\\
k_a^i \equiv 4g_{aa_1a_2}F_{i_1i_2}^{a_1}F_{0i_3}^{a_2}\e^{i_1i_2i_3i} = 0,
\end{cases}
\ee
where $\m=(0, i)$ and $i =1, \cdots, 2n$. 
Introduce the "$2nN\times 2nN$ matrix" $\Omega^{ij}_{ab}
\equiv 4\e^{ij i_1i_2}g_{abc}F^c_{i_1i_2}$ ($(b,j)$ as a collective index), 
we can rewrite the equations of motion in the compact form: 
\be \label{eq:EoMCS}
\begin{cases}
	k_a = \Omega_{ab}^{ij}F_{ij}^b = 0 \\
	\Omega_{ab}^{ij} F^b_{0j} = 0
\end{cases}
\ee

A simple identity 
\be \label{eq:appID}
	\d^i_{[k}g^{abc} \e^{i\ell mn} F^b_{j\ell}F^c_{mn]} = 0,
\quad\Rightarrow\quad 
	\Omega_{ab}^{ij} F^b_{kj} = \d^i_k k_a
\ee
shows that on the constraint surface $k_a = 0$, 
$(v_k)^b_j \equiv F_{kj}^b$ gives $2n$ null vectors to $\Omega_{ab}^{ij}$. 
The non-invertibility of $\Omega$ is due to the existence of symmetry. 
In this case, the $2n$ null vectors $F_{kj}^b$ generates the spatial 
diffeomorphism. 
In fact under diffeomorphism $\d x^\m = \eta^\m$  of spacetime, 
the Chern-Simons theory is invariant 
with $\d_{\eta}A_\m^a = \cL_{\eta}A_\m^a$, or the improved diffeomorphism 
\be
	\d_{\eta}A_\m^a = -\e^\n F_{\m\n}^a.
\ee 

In general, the rank of $\Omega$ depends on the 
properties of the invariant tensor $g_{abc}$, 
and the phase space location of the system. For example, at $F_{\m\n}^a=0$, 
$\Omega_{ab}^{ij} = 0$ and has zero rank. 
In \cite{HT1,HT2}, a \textit{generic condition} on $g_{abc}$ was 
introduced. $g_{abc}$ is said to be 
\textit{generic} if there exists solution $F_{ij}^a$ on the surface $k_a = 0$ 
such that: 
\begin{enumerate}
\item[(a)] The matrix $F_{kj}^b$ ($(b,j)$ as row and $k$ as column index) 
has the maximum rank 
	$2n$ such that $\xi^k F_{kj}^b=0$ implies $\xi^k = 0$, i.e. 
	the $2n$ null vectors $(v_k)^b_j \equiv F_{kj}^b$ of $\Omega_{ab}^{ij}$ 
are linearly independent.  
\item[(b)] The matrix $\Omega_{ab}^{ij}$ has maximum rank compatible
  with (a), 
i.e. 
	$\Omega_{ab}^{ij}$ has no other null vectors except $(v_k)^b_j$ and so has 
	rank $2nN-2n$
\end{enumerate}
We remark that the presence of the null vectors of $\Omega$ on the surface 
$k_a=0$ is due to
the presence of spatial diffeomorphism $\d x^i = \eta^i$, $i=1,2,3,4$. 
(under generic condition assumption, temporal diffeomorphism is not 
independent). If there were no 
such diffeomorphism, we would not expect the existence of such null vectors. 

Now the equation of motion (\ref{eq:EoMCS}) 
together with the generic condition implies 
$F_{0j}^b = N^k F_{kj}^b$ for arbitrary $2n$ fields $N^k$, or 
\be \label{eq:EoM+G}
	\dot{A}_i^a = D_i A_0^a + N^k F_{ki}^a
\ee
Since (\ref{eq:EoM+G}) is invariant under 
\begin{itemize}
\item[(a)] Standard gauge transformation ($\cN$ dimensional) : 
	\be\label{eq:appstdG}
		\d A_i^a = - D_i \l^a,\quad 
\d_\l A_0^a = -\dot{\l}^a - [\l,A_0]^a,\quad \d_\l N^k = 0
	\ee
\item[(b)] Spatial diffeomorphism ($2n$ dimensional) :
	\be\label{eq:appspatialD}
		\d_\xi A_i^a = -\xi^j F_{ij}^a,\quad 
\d_\xi A_0^a = -\xi^j F_{0j}^a, 
		\quad \d_\xi N^k = \dot{\xi}^k + [\xi,N]^k
	\ee
	where $[\xi,N]^k$ is the Lie bracket of the vectors $\xi$ and $N$,
\end{itemize}
we can use the above symmetries to go to the the time gauge 
\be
	A_0 = 0,\qquad N^k = 0.
\ee
In this case, the equation of motion is equivalent to 
\be
	k_a=0,\qquad A_i^a = \text{time independent}.
\ee
In addition to the $\cN$ constraints $k_a=0$, the  
$2n\cN$ functions $A^a_i(x_i)$ 
are subjected to  the residual symmetry of the time gauge, these are
$\cN$ time-independent 
gauge symmetry (\ref{eq:appstdG}) as well as the $2n$ time-independent 
spatial diffeomorphism (\ref{eq:appspatialD}), therefore 
the number of arbitrary functions in the solution to the equation of 
motion of Lagrange formulation is 
$
	2n\cN - \cN - (\cN+2n) = 2(n\cN-\cN-n). 
$ 
The local degrees of freedom is simply the half of it, therefore 
\be
	\text{no. of local degrees of freedom of pure CS} ~=~ n\cN - \cN - n
\ee
with $n>1$. 
In 5d,  this would be $\cN-2$. 
We remark that the above analysis holds only for the non-abelian case. 
For the counting of local degrees of freedom in the abelian case, 
see \cite{HT1,HT2}. 

\subsection{Chern-Simons theory coupled to conserved current}

For the case that the Chern-Simons theory is coupled to a conserved current
$J^\l$ ($D_\l J^\l =0$):
\be 
	S = \int d^5 x \; \tr\;  A_\m J_\m
+ S_{\text{CS}} ,
\ee
the equation of motion of $A_\l$ is 
\be
g_{aa_1a_2}F_{\m\n}^{a_1}F_{\l\s}^{a_2}\e^{\m\n\l\s\r} = cJ^a_\r
\ee
where $c$ is some constant.
In terms of the matrix 
$ \Omega_{ab}^{ij}\equiv \e^{iji_1i_2}g_{abc}F_{i_1i_2}^c$, 
the equation of motion can be written as 
\be \label{eq:appEoMCS2F}
\begin{cases} 
	\Omega_{ab}^{ij} F_{ij}^b = cJ_0^a \\
	4\Omega_{ab}^{ij} F_{0j}^b = c J_i^a
\end{cases}
\ee
Generically, $J^a_i \neq 0$, this means that (\ref{eq:appID}) 
can no longer be used to reduce the rank 
of $\Omega$, so we have full rank $2n\cN$ for $\Omega$ generically, i.e. 
$\Omega$ is invertible. 

Now in the gauge $A_0^a = 0$, the second line of the equation of 
motion (\ref{eq:appEoMCS2F}) simply provides a
first order partial differential equation in time: 
\be\label{eq:appCS2Fic}
	\del_0 A_j^b = c(\Omega^{-1})^{ab}_{ji} J^a_i .
\ee
As for the first equation of motion of (\ref{eq:appEoMCS2F}), it is indeed 
time-independent since 
\bea
	&&\del_0 (\Omega_{ab}^{ij}F_{ij}^b - cJ_0^a) 
	~=~ \Big( 2g_{abc}\del_0 F_{k\ell}^b F_{ij}^c \e^{ijk\ell} 
- c\del_0 J_0^a \Big) \nn\\
	&=& D_k [4g_{abc} F^b_{ij} F^c_{0\ell} \e^{ijk\ell}]  - c  D_i J_i^a
	~=~ c D_k J_k^a - c D_k J_k^a ~=~ 0
\eea
As a result, \eq{eq:appEoMCS2F} simply provides a constraint on the initial
values $A_j^b(x_i,t=0)$.
Therefore, in the time gauge, 
$A_j^b(x_i,t)$ are determined by (\ref{eq:appCS2Fic}) up to the
initial 
conditions 
$A_j^b(x_i,t=0)$. 
Both the time-independent gauge transformation 
and the  time-independent constraints  \eq{eq:appEoMCS2F}
remove $\cN$ independent initial conditions, 
so we have local degrees of freedom 
\be
	\frac{1}{2}(2n\cN - \cN - \cN) = (n-1)\cN 
\ee
In 5d, it's $\cN$. 

\section*{Acknowledgements}

It is a pleasure to thank Paul Heslop, Pei-Ming Ho,
Douglas Smith and Dimitri Sorokin for useful discussions and comments.
The work is partially supported by a STFC Consolidated Grant ST/J000426/1.

%c3
%c4 We also thank the referee for suggestions on clarification.
%c4 The research has been supported by  STFC.


\begin{thebibliography}{99}



\bm{zero}
G.~W.~Gibbons and P.~K.~Townsend,
  ``Vacuum interpolation in supergravity via super p-branes,''
  Phys.\ Rev.\ Lett.\  {\bf 71} (1993) 3754
  [hep-th/9307049].
  %%CITATION = HEP-TH/9307049;%%
\\
 A.~Strominger,
  ``Open p-branes,''
  Phys.\ Lett.\ B {\bf 383} (1996) 44
  [hep-th/9512059].
  %%CITATION = HEP-TH/9512059;%%
\\
 D.~M.~Kaplan and J.~Michelson,
  ``Zero modes for the D = 11 membrane and five-brane,''
  Phys.\ Rev.\ D {\bf 53} (1996) 3474
  [hep-th/9510053].
  %%CITATION = HEP-TH/9510053;%%
\\
 E.~Witten,
  ``Five-brane effective action in M theory,''
  J.\ Geom.\ Phys.\  {\bf 22} (1997) 103
  [hep-th/9610234].
  %%CITATION = HEP-TH/9610234;%%  

\bm{howe}
P.~S.~Howe and E.~Sezgin,
  ``D = 11, p = 5,''
  Phys.\ Lett.\ B {\bf 394} (1997) 62
  [hep-th/9611008].
  %%CITATION = HEP-TH/9611008;%%
\\
 P.~S.~Howe, E.~Sezgin and P.~C.~West,
  ``Covariant field equations of the M theory five-brane,''
  Phys.\ Lett.\ B {\bf 399} (1997) 49
  [hep-th/9702008].
  %%CITATION = HEP-TH/9702008;%%


\bm{PS} 
M.~Perry, J.~H.~Schwarz,
  ``Interacting chiral gauge fields in six-dimensions and Born-Infeld
theory,''
  Nucl.\ Phys.\  {\bf B489 } (1997)  47-64.
  [hep-th/9611065].

\bm{schw1}
 M.~Aganagic, J.~Park, C.~Popescu, J.~H.~Schwarz,
  ``World volume action of the M theory five-brane,''
  Nucl.\ Phys.\  {\bf B496 } (1997)  191-214.
  [hep-th/9701166].

\bm{pst}
 P.~Pasti, D.~P.~Sorokin, M.~Tonin,
  ``On Lorentz invariant actions for chiral p forms,''
  Phys.\ Rev.\  {\bf D55 } (1997)  6292-6298.
  [hep-th/9611100].
\\
P.~Pasti, D.~P.~Sorokin, M.~Tonin,
  ``Covariant action for a D = 11 five-brane with the chiral field,''
  Phys.\ Lett.\  {\bf B398 } (1997)  41-46.
  [hep-th/9701037].
\\
I.~A.~Bandos, K.~Lechner, A.~Nurmagambetov, P.~Pasti, D.~P.~Sorokin, M.~Tonin,
  ``Covariant action for the superfive-brane of M theory,''
  Phys.\ Rev.\ Lett.\  {\bf 78 } (1997)  4332-4334.
  [hep-th/9701149].
\\
I.~A.~Bandos, K.~Lechner, A.~Nurmagambetov, P.~Pasti, D.~P.~Sorokin 
and M.~Tonin,
  ``On the equivalence of different formulations of the M theory five-brane,''
  Phys.\ Lett.\ B {\bf 408} (1997) 135
  [hep-th/9703127].
  %%CITATION = HEP-TH/9703127;%%

\bm{nilsson}
 M.~Cederwall, B.~E.~W.~Nilsson and P.~Sundell,
  ``An Action for the superfive-brane in $D = 11$ supergravity,''
  JHEP {\bf 9804} (1998) 007
  [hep-th/9712059].
  %%CITATION = HEP-TH/9712059;%%

\bm{no-go}
 M.~Henneaux and B.~Knaepen,
  ``All consistent interactions for exterior form gauge fields,''
  Phys.\ Rev.\ D {\bf 56} (1997) 6076
  [hep-th/9706119].
  %%CITATION = HEP-TH/9706119;%%
\\
M.~Henneaux,
  ``Uniqueness of the Freedman-Townsend interaction vertex for 
two form gauge fields,''
  Phys.\ Lett.\ B {\bf 368} (1996) 83
  [hep-th/9511145].
  %%CITATION = HEP-TH/9511145;%%
\\
 X.~Bekaert, M.~Henneaux and A.~Sevrin,
  ``Deformations of chiral two forms in six-dimensions,''
  Phys.\ Lett.\ B {\bf 468} (1999) 228
  [hep-th/9909094].
  %%CITATION = HEP-TH/9909094;%%
\\
 M.~Henneaux and B.~Knaepen,
  ``A Theorem on first order interaction vertices for free p form gauge fields,''
  Int.\ J.\ Mod.\ Phys.\ A {\bf 15} (2000) 3535
  [hep-th/9912052].
  %%CITATION = HEP-TH/9912052;%%
\\
  R.~I.~Nepomechie,
  ``Approaches To A Nonabelian Antisymmetric Tensor Gauge Field Theory,''
  Nucl.\ Phys.\ B {\bf 212} (1983) 301.
  %%CITATION = NUPHA,B212,301;%%
\\
X.~Bekaert, M.~Henneaux and A.~Sevrin,
  ``Chiral forms and their deformations,''
  Commun.\ Math.\ Phys.\  {\bf 224} (2001) 683
  [hep-th/0004049].
  %%CITATION = HEP-TH/0004049;%%
\\
X.~Bekaert and S.~Cucu,
  ``Deformations of duality symmetric theories,''
  Nucl.\ Phys.\ B {\bf 610} (2001) 433
  [hep-th/0104048].
  %%CITATION = HEP-TH/0104048;%%
\\
 C.~-H.~Chen, P.~-M.~Ho and T.~Takimi,
  ``A No-Go Theorem for M5-brane Theory,''
  JHEP {\bf 1003} (2010) 104
  [arXiv:1001.3244 [hep-th]].
  %%CITATION = ARXIV:1001.3244;%%


\bibitem{CS2}
C.~S.~Chu and D.~J.~Smith,
  ``Multiple Self-Dual Strings on M5-Branes,''\\
  JHEP {\bf 1001} (2010) 001
  [arXiv:0909.2333 [hep-th]].


\bibitem{ABJM}
 O.~Aharony, O.~Bergman, D.~L.~Jafferis and J.~Maldacena,
  ``N=6 superconformal Chern-Simons-matter theories, M2-branes and their
  gravity duals,''
  JHEP {\bf 0810} (2008) 091
  [arXiv:0806.1218 [hep-th]].
  %%CITATION = JHEPA,0810,091;%%


\bm{chu}
 C.~-S.~Chu,
  ``A Theory of Non-Abelian Tensor Gauge Field with 
Non-Abelian Gauge Symmetry G x G,''
  arXiv:1108.5131 [hep-th].
  %%CITATION = ARXIV:1108.5131;%%

\bm{HT}
M.~Henneaux and C.~Teitelboim,
  ``Dynamics Of Chiral (selfdual) P Forms,''
  Phys.\ Lett.\ B {\bf 206} (1988) 650.
  %%CITATION = PHLTA,B206,650;%%

\bibitem{BH} 
  C.~Bunster and M.~Henneaux,
  ``The Action for Twisted Self-Duality,''
  Phys.\ Rev.\ D {\bf 83}, 125015 (2011)
  [arXiv:1103.3621 [hep-th]].
  %%CITATION = ARXIV:1103.3621;%%

\bibitem{dou}
M.~R.~Douglas,
  ``On D=5 super Yang-Mills theory and (2,0) theory,''
  JHEP {\bf 1102 } (2011)  011.
  [arXiv:1012.2880 [hep-th]].

\bibitem{lam2}
  N.~Lambert, C.~Papageorgakis, M.~Schmidt-Sommerfeld,
  ``M5-Branes, D4-Branes and Quantum 5D super-Yang-Mills,''
  JHEP {\bf 1101 } (2011)  083.
  [arXiv:1012.2882 [hep-th]].


\bibitem{lam1}
 N.~Lambert, C.~Papageorgakis,
  ``Nonabelian (2,0) Tensor Multiplets and 3-algebras,''
  JHEP {\bf 1008 } (2010)  083.
  [arXiv:1007.2982 [hep-th]].


\bibitem{ho}
P.~-M.~Ho, K.~-W.~Huang, Y.~Matsuo,
  ``A Non-Abelian Self-Dual Gauge Theory in 5+1 Dimensions,''
  JHEP {\bf 1107 } (2011)  021.
  [arXiv:1104.4040 [hep-th]].

\bibitem{sezgin}
  H.~Samtleben, E.~Sezgin, R.~Wimmer,
  ``(1,0) superconformal models in six dimensions,''  
  [arXiv:1108.4060 [hep-th]].

\bm{qg}
 C.~-S.~Chu and D.~J.~Smith,
  ``Towards the Quantum 
Geometry of the M5-brane in a Constant C-Field from Multiple Membranes,''
  JHEP {\bf 0904} (2009) 097
  [arXiv:0901.1847 [hep-th]].
  %%CITATION = ARXIV:0901.1847;%%
\\
  J.~DeBellis, C.~Saemann, R.~J.~Szabo,
  %``Quantized Nambu-Poisson Manifolds and n-Lie Algebras,''
  J.\ Math.\ Phys.\  {\bf 51 } (2010)  122303.
  [arXiv:1001.3275 [hep-th]].
\\
  J.~DeBellis, C.~Samann, R.~J.~Szabo,
  ``Quantized Nambu-Poisson Manifolds in a 3-Lie Algebra Reduced Model,''
  JHEP {\bf 1104 } (2011)  075.
  [arXiv:1012.2236 [hep-th]]

\bm{CG}
C.~-S.~Chu, G.~S.~Sehmbi,
``D1-Strings in Large RR 3-Form Flux, Quantum Nambu Geometry
and M5-Branes in C-Field,''
[arXiv:1110.2687 [hep-th]].

\bm{qg2}
 P.~-M.~Ho and Y.~Matsuo,
  %``M5 from M2,''
  JHEP {\bf 0806} (2008) 105
  [arXiv:0804.3629 [hep-th]].
  %%CITATION = ARXIV:0804.3629;%%
\\
P.~-M.~Ho, Y.~Imamura, Y.~Matsuo and S.~Shiba,
  ``M5-brane in three-form flux and multiple M2-branes,''
  JHEP {\bf 0808} (2008) 014
  [arXiv:0805.2898 [hep-th]].
  %%CITATION = ARXIV:0805.2898;%%
\\
 P.~Pasti, I.~Samsonov, D.~Sorokin and M.~Tonin,
  ``BLG-motivated Lagrangian formulation for the chiral two-form gauge field
in D=6 and M5-branes,''
  Phys.\ Rev.\ D {\bf 80} (2009) 086008
  [arXiv:0907.4596 [hep-th]].
\\
K.~Furuuchi,
  ``Non-Linearly Extended Self-Dual Relations From 
The Nambu-Bracket Description Of M5-Brane In A Constant C-Field Background,''
  JHEP {\bf 1003} (2010) 127
  [arXiv:1001.2300 [hep-th]].
  %%CITATION = ARXIV:1001.2300;%%

\bm{amp}
 B.~Czech, Y.~-t.~Huang and M.~Rozali,
  %``Amplitudes for Multiple M5 Branes,''
  arXiv:1110.2791 [hep-th].
  %%CITATION = ARXIV:1110.2791;%%


\bm{lee}
S.~Bolognesi and K.~Lee,
  ``1/4 BPS String Junctions and $N^3$ Problem in 6-dim (2,0) 
Superconformal Theories,''
  Phys.\ Rev.\ D {\bf 84} (2011) 126018
  [arXiv:1105.5073 [hep-th]].
  %%CITATION = ARXIV:1105.5073;%%
\\
H.~-C.~Kim, S.~Kim, E.~Koh, K.~Lee and S.~Lee,
  ``On instantons as Kaluza-Klein modes of M5-branes,''
  JHEP {\bf 1112} (2011) 031
  [arXiv:1110.2175 [hep-th]].
  %%CITATION = ARXIV:1110.2175;%%

\bm{others}
S.~Terashima and F.~Yagi,
  ``On Effective Action of Multiple M5-branes and ABJM Action,''
  JHEP {\bf 1103} (2011) 036
  [arXiv:1012.3961 [hep-th]].
  %%CITATION = ARXIV:1012.3961;%%
\\
H.~Singh,
  ``Super-Yang-Mills and M5-branes,''
  JHEP {\bf 1108}, 136 (2011)
  [arXiv:1107.3408].
\\
 Y.~Tachikawa,
  ``On S-duality of 5d super Yang-Mills on $S^1$,''
  JHEP {\bf 1111} (2011) 123
  [arXiv:1110.0531 [hep-th]].
  %%CITATION = ARXIV:1110.0531;%%
\\
C.~Saemann and M.~Wolf,
  ``On Twistors and Conformal Field Theories from Six Dimensions,''
  arXiv:1111.2539 [hep-th].
  %%CITATION = ARXIV:1111.2539;%%  
\\
 A.~Gustavsson,
  ``M5 brane on $R^{1,2} \times S^3$,''
  JHEP {\bf 1201} (2012) 057
  [arXiv:1111.5392 [hep-th]].
  %%CITATION = ARXIV:1111.5392;%%
\\
N.~Lambert, H.~Nastase and C.~Papageorgakis,
  ``5D Yang-Mills instantons from ABJM Monopoles,''
  arXiv:1111.5619 [hep-th].
  %%CITATION = ARXIV:1111.5619;%%
\\
 A.~Gustavsson,
  ``A preliminary test of Abelian D4-M5 duality,''
  Phys.\ Lett.\ B {\bf 706} (2011) 225
  [arXiv:1111.6339 [hep-th]].
  %%CITATION = ARXIV:1111.6339;%%

\bm{r1}
 D.~S.~Berman,
  ``M-theory branes and their interactions,''
  Phys.\ Rept.\  {\bf 456} (2008) 89
  [arXiv:0710.1707 [hep-th]].
  %%CITATION = ARXIV:0710.1707;%%
\\
P.~Arvidsson,
  ``Superconformal Theories in Six Dimensions,''
  hep-th/0608014.
  %%CITATION = HEP-TH/0608014;%%


\bm{HT1}
 M.~Banados, L.~J.~Garay and M.~Henneaux,
  ``The Local degrees of freedom of higher dimensional pure Chern-Simons theories,''
  Phys.\ Rev.\ D {\bf 53} (1996) 593
  [hep-th/9506187].
  %%CITATION = HEP-TH/9506187;%%

\bm{HT2}
  M.~Banados, L.~J.~Garay and M.~Henneaux,
  ``The Dynamical structure of higher dimensional Chern-Simons theory,''
  Nucl.\ Phys.\ B {\bf 476} (1996) 611
  [hep-th/9605159].
  %%CITATION = HEP-TH/9605159;%%

\bm{HT3}
 O.~Miskovic, R.~Troncoso and J.~Zanelli,
  ``Canonical sectors of five-dimensional Chern-Simons theories,''
  Phys.\ Lett.\ B {\bf 615} (2005) 277
  [hep-th/0504055].
  %%CITATION = HEP-TH/0504055;%%


\bm{GSW}
 M.~B.~Green, J.~H.~Schwarz and E.~Witten,
  ``Superstring Theory. Vol. 1: Introduction,'' Chapter 2.3, 
  Cambridge, Uk: Univ. Pr. ( 1987).

\bm{dbi}
  A.~A.~Tseytlin,
  ``On nonAbelian generalization of Born-Infeld action in string theory,''
  Nucl.\ Phys.\ B {\bf 501} (1997) 41
  [hep-th/9701125].
  %%CITATION = HEP-TH/9701125;%%
\\
  A.~A.~Tseytlin,
  ``Born-Infeld action, supersymmetry and string theory,''
  In *Shifman, M.A. (ed.): The many faces of the superworld* 417-452
  [hep-th/9908105].
  %%CITATION = HEP-TH/9908105;%%
\\
  P.~Koerber and A.~Sevrin,
  ``The NonAbelian D-brane effective action through order alpha-prime**4,''
  JHEP {\bf 0210} (2002) 046
  [hep-th/0208044].
  %%CITATION = HEP-TH/0208044;%%

\bm{wit}
E.~Witten,
  ``Conformal Field Theory In Four And Six Dimensions,''
  arXiv:0712.0157 [math.RT].
  %%CITATION = ARXIV:0712.0157;%%

\bm{dual}
I.~A.~Bandos, N.~Berkovits and D.~P.~Sorokin,
 ``Duality symmetric eleven-dimensional supergravity and its coupling to
M-branes,''
  Nucl.\ Phys.\ B {\bf 522} (1998) 214
  [hep-th/9711055].


\bm{5dcs1}
 N.~Seiberg,
  ``Five-dimensional SUSY field theories, nontrivial fixed points and string dynamics,''
  Phys.\ Lett.\ B {\bf 388} (1996) 753
  [hep-th/9608111].
  %%CITATION = HEP-TH/9608111;%%
\\
K.~A.~Intriligator, D.~R.~Morrison and N.~Seiberg,
  ``Five-dimensional supersymmetric gauge theories and degenerations of Calabi-Yau spaces,''
  Nucl.\ Phys.\ B {\bf 497} (1997) 56
  [hep-th/9702198].
  %%CITATION = HEP-TH/9702198;%%
\\
 M.~Aganagic, M.~Marino and C.~Vafa,
  ``All loop topological string amplitudes from Chern-Simons theory,''
  Commun.\ Math.\ Phys.\  {\bf 247} (2004) 467
  [hep-th/0206164].
  %%CITATION = HEP-TH/0206164;%%

\bm{5dcs2}
 A.~Iqbal and A.~-K.~Kashani-Poor,
  ``SU(N) geometries and topological string amplitudes,''
  Adv.\ Theor.\ Math.\ Phys.\  {\bf 10} (2006) 1
  [hep-th/0306032].
  %%CITATION = HEP-TH/0306032;%%
\\
  Y.~Tachikawa,
  ``Five-dimensional Chern-Simons terms and Nekrasov's instanton counting,''
  JHEP {\bf 0402} (2004) 050
  [hep-th/0401184].
  %%CITATION = HEP-TH/0401184;%%

\bm{chub}
C.~-S.~Chu, work in progress.

\bibitem{BLG}
  J.~Bagger and N.~Lambert,
  ``Gauge Symmetry and Supersymmetry of Multiple M2-Branes,''
  Phys.\ Rev.\  D {\bf 77} (2008) 065008
  [arXiv:0711.0955 [hep-th]].
  %%CITATION = PHRVA,D77,065008;%%
\\
  J.~Bagger and N.~Lambert,
  ``Comments On Multiple M2-branes,''
  JHEP {\bf 0802} (2008) 105
  [arXiv:0712.3738 [hep-th]].
  %%CITATION = JHEPA,0802,105;%%
\\
  A.~Gustavsson,
  ``Algebraic structures on parallel M2-branes,''
  arXiv:0709.1260 [hep-th].
  %%CITATION = ARXIV:0709.1260;%%

\bibitem{BFSS}
  T.~Banks, W.~Fischler, S.~H.~Shenker and L.~Susskind,
  ``M theory as a matrix model: A Conjecture,''
  Phys.\ Rev.\ D {\bf 55} (1997) 5112
  [hep-th/9610043].
  %%CITATION = HEP-TH/9610043;%%

\bm{IKKT}
  N.~Ishibashi, H.~Kawai, Y.~Kitazawa and A.~Tsuchiya,
  ``A Large N reduced model as superstring,''
  Nucl.\ Phys.\ B {\bf 498} (1997) 467
  [hep-th/9612115].
  %%CITATION = HEP-TH/9612115;%%

\bm{t1}
  W.~Taylor,
  ``M(atrix) theory: Matrix quantum mechanics as a fundamental theory,''
  Rev.\ Mod.\ Phys.\  {\bf 73} (2001) 419
  [hep-th/0101126].
  %%CITATION = HEP-TH/0101126;%%

\bm{t2}
M.~R.~Douglas,
  ``D-branes and matrix theory in curved space,''
  Nucl.\ Phys.\ Proc.\ Suppl.\  {\bf 68} (1998) 381
  [hep-th/9707228].
  %%CITATION = HEP-TH/9707228;%%
\\
 M.~R.~Douglas, A.~Kato and H.~Ooguri,
  ``D-brane actions on Kahler manifolds,''
  Adv.\ Theor.\ Math.\ Phys.\  {\bf 1} (1998) 237
  [hep-th/9708012].
  %%CITATION = HEP-TH/9708012;%%

\bm{t3}
  M.~Hanada, H.~Kawai and Y.~Kimura,
  ``Describing curved spaces by matrices,''
  Prog.\ Theor.\ Phys.\  {\bf 114} (2006) 1295
  [hep-th/0508211].
  %%CITATION = HEP-TH/0508211;%%

\bm{m5-S}
 I.~R.~Klebanov and A.~A.~Tseytlin,
  ``Entropy of near extremal black p-branes,''
  Nucl.\ Phys.\ B {\bf 475} (1996) 164
  [hep-th/9604089].
  %%CITATION = HEP-TH/9604089;%%



\end{thebibliography}
\end{document}